\documentclass[aps,superscriptaddress,showpacs,nofootinbib,eqsecnum,prd,notitlepage]{revtex4} 

\usepackage{amssymb}   
\usepackage{amsmath}   
\usepackage{natbib}
\usepackage{epsfig}   
\usepackage{graphicx}   
\usepackage{dcolumn}
\usepackage{bm}  
\usepackage[english]{babel}
\usepackage[latin1]{inputenc}
\usepackage{eucal} 
\usepackage{hyperref}
\usepackage{verbatim}
\usepackage{latexsym}
\usepackage{xcolor}

\usepackage{epstopdf}
\begin{document}

\title{
Eternal inflation in a dissipative and radiation environment:
Heated demise of eternity 
}

\author{Gustavo S. Vicente} \email{gsvicente@uerj.br, 
gustavosvicente@gmail.com}
\affiliation{Departamento de F\'{\i}sica Te\'orica, Universidade do
  Estado do Rio de Janeiro, 20550-013 Rio de Janeiro, Rio de Janeiro, Brazil}
  
\author{Leandro A. da Silva} \email{leandro.silva@ufabc.edu.br}
\affiliation{Center of Mathematics, Computation and Cognition, UFABC, 
09210-580 Santo Andr\'e, S\~ao Paulo, Brazil}  
  
\author{Rudnei O.  Ramos} \email{rudnei@uerj.br}
\affiliation{Departamento de F\'{\i}sica Te\'orica, Universidade do
  Estado do Rio de Janeiro, 20550-013 Rio de Janeiro, RJ, Brazil}
  
\begin{abstract}

Eternal inflation is studied in the context of warm inflation. We
focus on different tools to analyze the effects of dissipation and the
presence of a thermal radiation bath on the fluctuation-dominated
regime, for which the self-reproduction of Hubble regions can take place.
The tools we explore are the {\it threshold inflaton field} and
{\it threshold number of {\it e}-folds} necessary to establish a
self-reproduction regime and the {\it counting of Hubble regions},
using generalized conditions for the occurrence of a
fluctuation-dominated regime. We obtain the functional dependence of
these quantities on the dissipation and temperature. A
Sturm-Liouville analysis of the Fokker-Planck equation for the
probability of having eternal inflation and an analysis for the
probability of having eternal points are performed. We have
considered the representative cases of inflation models with monomial
potentials of the form of chaotic and hilltop ones. Our results show
that warm inflation tends to initially favor the onset of a
self-reproduction regime for smaller values of the dissipation. As
the dissipation increases, it becomes harder than in cold inflation
(i.e., in the absence of dissipation) to achieve a self-reproduction
regime  for both types of models analyzed. The results are interpreted
and explicit analytical expressions are given whenever that is possible.

\end{abstract}


\maketitle

\section{Introduction}
\label{sec1}

One of the most peculiar consequences of inflation is the possibility
of leading to a self-reproduction regime (SRR) of inflating
Hubble regions (H regions),  a phenomenon that became known as eternal
inflation (for reviews, see, e.g.,
Refs.~\cite{Guth2007,winitzki2009eternal,guth2000inflation}).  In
eternal inflation the dynamics of the Universe during the inflationary
phase is considered in a global perspective and refers to a
semi-infinity (past finite, future infinity) mechanism of
self-reproduction of causally disconnected
H regions~\cite{Vilenkin:1983xq,Guth:1985ya,Linde:1986fc,Linde:1986fd}.
Looking at the spacetime structure as a whole, the distribution of
H regions resembles  much like that of a bubble foam. This scenario
seems to be a generic feature present in several models of
inflation. In recent years, eternal inflation attracted renewed
attention due to several factors. One of them is the seemingly
intrinsic connection between eternal inflation and extra dimensions
theories, like string theory, and its multitude of possible solutions
describing different false vacua,  each one yielding its own
low-energy
constants~\cite{vilenkin2007measure,Guth2007,garriga2006probabilities},
leading to what has been known by the ``multiverse'', when combined with
eternal inflation.

Collisions between pocket universes could upset in some level the
homogeneity and isotropy of the bubble we live in and, therefore, lead
to some detectable signature in the cosmic microwave background
radiation (CMBR). This has then led to different proposals to test
eternal
inflation~\cite{zhang2015testing,wainwright2014simulating,wainwright2014simulating2,
  feeney2011first,feeney2011first2,aguirre2011status}.  Eternal
inflation has also been recently studied in experiments involving
analogue systems in condensed matter. {}For example, in
Ref.~\cite{smolyaninov2013experimental}, an analogue model using
magnetic particles in a cobalt-based ferrofluid system has been used
to show that thermal fluctuations are capable of generating $2 +
1$-dimensional Minkowski-like regions inside a larger metamaterial
that plays the role of the background of the multiverse.  {}From the
model building point of view, a relatively recent trend is to look for
new models, or specific regimes in known models, where SRR may be
suppressed. If eternal inflation does not take place, then its typical
conceptual and predictive problems could be avoided, thus allowing the
return of a more simple picture of universe evolution. {}For example,
in Ref.~\cite{mukhanov2014inflation}, it is discussed as an extension of
a cold inflationary scenario where some requirements are established
such that a SRR could be suppressed.  In
Ref.~\cite{kinney2014negative}, the authors discuss the possibility of
preventing a SRR given a negative running of the scalar spectral index
on superhorizon scales, motivated by earlier results from
Planck~\cite{Planck2013} and by the BICEP2 experiment~\cite{BICEP2}.
In a more recent work~\cite{brandenberger2015can}, it is discussed how
backreaction effects impact on the stochastic growth of the inflaton
field. The authors in Ref.~\cite{brandenberger2015can} have concluded
that for a power-law and Starobinsky inflation, the strength of the
backreaction is too weak to  avoid eternal inflation, while  in cyclic
Ekpyrotic scenarios, the SRR could be prevented.

In this work, we want to study and then establish the conditions for
the presence of a  SRR under the framework of the warm inflation
picture~\cite{Berera:1995ie}.  In the standard inflationary picture,
usually known as \textit{cold inflation}, it is typically assumed that
the couplings of the inflaton field to other field degrees of freedom
are negligible during inflation, becoming only relevant later on in
order to produce a successful preheating/reheating phase, leading to a
thermal radiation bath when the decay products of the inflaton field
thermalize.  In cold inflation, density fluctuations are mostly
sourced by {\it quantum fluctuations} of the inflaton field
\cite{lyth2009primordial}.  On the other case, in the warm inflation
picture, it may happen that the couplings among the various fields are
sufficiently strong to effectively generate and keep a
quasiequilibrium thermal radiation bath throughout the inflationary
phase. In this situation, the inflationary phase can be smoothly
connected to the radiation dominated epoch, without the need, {\it a
priori}, of a separate reheating period (for reviews, see, e.g.,
Refs.~\cite{WIreviews1,WIreviews2}).  In warm inflation, the primary
source of density fluctuations comes from {\it thermal fluctuations},
which originate in the radiation bath and are transferred to the
inflaton in the form of adiabatic curvature
perturbations~\cite{warmpert1,warmpert2}.  

We know from many recent
studies~\cite{ng1,ng2,Liu:2014ifa,LAS2013,bartrum2014importance,bastero2014cosmological,Bastero-Gil:2014raa}
that dissipation and stochastic noise effects are able to strongly
modify the inflationary dynamics. This in turn can lead to very
different predictions for observational quantities, like for the
tensor-to-scalar ratio, the spectral index, and nongaussianities, when
compared to the cold inflation case.  Thus, it is natural to expect
that those intrinsic dynamic changes in warm inflation due to
dissipation and the presence of the thermal radiation bath can and
should potentially affect the predictions concerning eternal inflation
as well. 

Warm inflation has been studied only from the ``local'' perspective,
where only the space-time region causally accessible from one
worldline is described. On the other hand, the insertion of a warm
inflation features in the context of a ``global'' picture, where the
eternal inflation  description becomes relevant, has been neglected so
far.  The different predictions of cold and warm inflation concerning
the conditions for the establishment of a SRR regime could result in
one more tool to select the most realistic model given appropriate
observational constraints. The main question we aim to address in this
paper is how the presence of dissipation, stochastic  noise, and a
thermal bath generated through dissipative effects during warm
inflation will affect the global structure of the inflationary
universe. {}For this, we develop a generalized eternal inflation model
of random walk type in the context of warm inflation and use standard
tools like the Sturm-Liouville analysis (SLA) of the {}Fokker-Planck
equation associated with the random process and the analysis of the
presence of eternal points, which allow us to verify the presence of a
fluctuation-dominated range (FDR). In addition, we introduce the
analysis of the threshold value of the inflaton field and the
threshold number of {\it e}-folds  for the existence of a FDR and the
counting of Hubble regions produced during the global (warm)
inflationary evolution in order to assess how warm inflation modifies
typical measures of eternal inflation.   In this work, we do not intend
to address the known  conceptual and prediction issues usually
associated with eternal inflation (for a recent discussion of these
issues and for the different point of views on these matters, see,
e.g., Refs.~\cite{Ijjas:2013vea,Guth:2013sya,Linde:2014nna}). 

This paper is organized as follows. In Sec.~\ref{sec2}, we briefly
review the basics of random walk eternal inflation in the cold
inflation context. The different ways of characterizing eternal
inflation, and those we will be using in this work are also reviewed.
In Sec. \ref{sec3}, the ideas of warm and eternal inflation  are
combined and a generalized model is described. The relevant results
are discussed  in Sec.~\ref{results} and, finally, our concluding
remarks are given in Sec. \ref{sec6}.  We also include two appendices
where we give some of the technical details used to derive our results
and also to explain the numerical analysis we have employed.
  
%%%%%%%%%%%%%%%%%%%%%%%%%%%%%%%%%%%%%%%%%%%%%%%%%%%%%%%%%%%%%%%%%%%%%%%%%%
\section{Characterizing Eternal inflation: a brief exposition } 
\label{sec2}

Eternal inflation refers to the property of the inflationary regime
having no end when we look at the spacetime structure as a whole. This
scenario is a generic feature present in several inflation models,
provided that certain conditions are met, as we will discuss
below. Mathematically, the formulation that allow us  to model eternal
inflation is mostly conveniently expressed in terms of the Starobinsky
stochastic inflation program, which describes the backreaction of the
short wavelength modes, which get frozen at the horizon crossing, into the
dynamics of the long wavelength inflaton
modes~\cite{original,Starobinsky1994}. In this context, the standard
equation of motion for the inflaton field $\varphi$ can be written as
a Langevin-like equation of the form~\cite{original}

\begin{equation} \label{eomEI}
\dot \varphi =  f(\varphi) + \sqrt{2D^{(2)}(\varphi)}\zeta \;,
\end{equation}
where $f(\varphi) \equiv -V_{,\varphi}/(3H(\varphi))$ and
$D^{(2)}(\varphi) = H^3(\varphi)/(8\pi^2)$ are, respectively, the
drift and diffusion coefficients, and $\zeta$ is a Gaussian noise term
that accounts for the quantum fluctuations of the inflaton field,
whose correlation function is given by $\langle \zeta(t) \zeta(t')
\rangle  = \delta(t-t')$.   In de Sitter spacetime, we can show that
the inflaton fluctuations grow linearly as a function of
time~\cite{PhysRevD.26.1231,Linde1982335,Starobinsky1982175},

\begin{equation} \label{coldfluc}
 \langle \varphi^2(t+\Delta t) \rangle  - \langle \varphi^2(t) \rangle
 \sim \frac{H^3}{4\pi^2}\Delta t    \;.
\end{equation}
It is assumed that when subhorizon modes cross the horizon ($\sim
H^{-1}$), they become classical quantities in a sufficiently small time
interval. Consequently, the large-scale dynamics for the inflaton can
be seen as a random walk with a typical stepsize $\sim H/(2\pi)$ in a
time interval $\sim H^{-1}$. Defining $\varphi_{d}(t)$ as the
deterministic dynamics for the inflaton field, we can distinguish
between two typical regimes: i) if $\dot \varphi_{d} H^{-1}$ dominates
over the fluctuations, the slow-roll evolution of the inflaton field
is essentially deterministic; ii) in the opposite case, when the
fluctuations dominate over the $\dot \varphi_{d} H^{-1}$ term, then
the inflaton dynamics can be treated as a random walk and we have a
FDR. In the FDR, random fluctuations of the inflaton field may advance
or delay the onset of the reheating phase in different regions, avoiding
global reheating.  Given a value of $\varphi$ that is nearly
homogeneous in a region of the order of magnitude of the horizon size
(known as \textit{Hubble region} or \textit{H region}) and has a value
that satisfies the FDR, this H region will expand, generating seeds
for new H regions, and this process goes on indefinitely towards
future. In a sense, one can say that a requirement for the presence of
a SRR is  that the inflationary dynamics goes through a FDR.

The fluctuations of the inflaton field are represented in
Eq.~(\ref{eomEI}) by the noise term
$\sqrt{2D^{(2)}(\varphi)}\zeta$, whereas the term $f(\varphi)$
represents the deterministic evolution.  Therefore, a FDR occurs when
the following condition is satisfied~\cite{winitzki2009eternal}: 

\begin{equation} \label{condition}
\frac{|f(\varphi)|}{H} \ll \sqrt{\frac{2D^{(2)}}{H}}\;.    
\end{equation}
More precisely, the time evolution of the inflaton field is strongly
nondeterministic while the diffusion term dominates over the drift
one.  We call Eq.~(\ref{condition}) the FDR condition.

The FDR condition provides the values of the inflaton field for which
a FDR is set, serving as a sufficient tool to look for the presence of
eternal inflation.  However, we will see in the following sections
that as we leave the cold inflation context  and generalize the FDR
condition to warm inflaton, it acquires a nontrivial dependence in the
thermal bath variables, and we need to introduce additional tools in
order to appreciate the presence of eternal inflation.  For the sake
of comparison, we perform these approaches for both cold and warm
inflation.

In the following subsections, we introduce the SLA of the Fokker-Planck
equation and the analysis of the presence of eternal points, which are
generic for both cold and warm inflation cases.

%%%%%%%%%%%%%%%%%%%%%%%%%%%%%%%%%%%%%%%%%%%%%%%%%%%%%%%%%%%%%%%%%%%%%%%%%%
\subsection{Fokker-Planck equation}
\label{FDRsection}

Statistical properties of $\varphi$ can be obtained through the
probability density function   $P(\varphi,t)d\varphi$. This is a
function that describes the probability of finding the inflaton field
at a value $\varphi$ at time $t$, where the values of $\varphi$ are
measured in a worldline randomly chosen at constant  ${\bf
  x}$ coordinates  in a single H region.  $P(\varphi,t)$ is known as
the \textit{comoving probability distribution} and satisfies the
following {}Fokker-Planck equation:

\begin{equation} \label{FPcold}
\frac{\partial P}{\partial t} = \frac{\partial }{\partial \varphi}
\left[-D^{(1)}(\varphi) P +  \frac{\partial }{\partial \varphi} \left(
  D^{(2)}(\varphi)P  \right) \right]\;. 
\end{equation}
However, when one is interested in the global perspective, one has to
consider the \textit{volume weighted distribution}, $P_V(\varphi,t)$,
where the volume contains many H regions.  The expression
$P_V(\varphi,t)d\varphi$ is defined as the physical three-dimensional
volume $\int \sqrt{-g} d^3 {\bf x}$ of regions having the value
$\varphi$ at time $t$. This distribution satisfies the equation

\begin{equation} 
\label{FPVcold0}
\frac{\partial P_V}{\partial t} = \frac{\partial }{\partial \varphi}
\left[-D^{(1)}(\varphi) P_V +  \frac{\partial }{\partial \varphi}
  \left( D^{(2)}(\varphi)P_V  \right) \right] + 3 H(\varphi) P_V\;, 
\end{equation}
where the fundamental difference in relation to Eq.~(\ref{FPcold}) is
the presence of the $3 H(\varphi) P_V$ term, which describes the
exponential growth of a three-dimensional volume in regions under
inflationary expansion. We can also write a {}Fokker-Planck equation
for the distribution $P_V(\varphi,t)$, normalized to unity,
$P_P(\varphi,t) \equiv P_V(\varphi,t)/\left\langle \exp{\left(3\int dt
  H \right)}\right\rangle$, but it is sufficient for our analysis to
work with $P_V$.

To completely specify the probability distribution function, one needs
to assume certain boundary conditions. Exit boundary conditions,
$\frac{\partial }{\partial \varphi} \left[  D^{(2)}(\varphi)P
  \right]_{\varphi = \varphi_c} = 0$ and $\frac{\partial }{\partial
  \varphi} \left[  D^{(2)}(\varphi)P_V  \right]_{\varphi = \varphi_c}
= 0$, and/or absorbing boundary condition, $P(\varphi_c) = 0$, are
typically imposed at the end of inflation (reheating boundary or
surface),  $\varphi_c = \varphi_{f}$, and at the beginning of
inflation, $\varphi_c = \varphi_{i}$, where $\varphi_{i}$ and
$\varphi_{f}$ are, respectively, the initial and final values for the
inflaton field.  In our analysis, we will adopt the It\^o ordering and
the proper-time parametrization (for discussions concerning gauge
dependence  and factor ordering issues, see, for example,
Ref.~\cite{Winitzki1996}).

A general overdamped Langevin equation of the form

\begin{equation} \label{langevin}
\dot \varphi = f(\varphi) + g_1(\varphi) \zeta_1(t) + g_2(\varphi)
\zeta_2(t)\;, 
\end{equation}
with noises $\zeta_1$ and $\zeta_2$ satisfying

\begin{align} 
& \langle  \zeta_i(t) \rangle = 0~, \nonumber \\  &\langle  \zeta_i(t)
  \zeta_i(t') \rangle = \delta(t-t')~,  \nonumber \\  &\langle
  \zeta_i(t)  \zeta_j(t') \rangle = \theta \delta(t-t') \;,
\end{align}
possesses an associated Fokker-Planck equation (following the It\^o
prescription) given by

\begin{align} \label{warmFP}
\frac{\partial }{\partial t} P(\varphi,t) &=  -\frac{\partial
}{\partial \varphi} \left[D^{(1)}  P(\varphi,t) \right] +
\frac{\partial^2 }{\partial \varphi^2}  \left[D^{(2)}  P(\varphi,t)
  \right] \nonumber \\ & = - \frac{\partial }{\partial \varphi}
\left\{  D^{(1)} P(\varphi,t)  -  \frac{\partial }{\partial \varphi}
\left[D^{(2)}  P(\varphi,t) \right]\right\} \;.
\end{align}
The drift and diffusion coefficients are given, respectively, by
\cite{madureira1996giant}

\begin{align} \label{coeffs_FP}
& D^{(1)} =  f(\varphi) \;, \nonumber \\ &  D^{(2)} =
  \frac{g_1(\varphi)^2}{2} + \theta g_1(\varphi) g_2(\varphi) +
  \frac{g_2(\varphi)^2}{2} \;.
\end{align}

%%%%%%%%%%%%%%%%%%%%%%%%%%%%%%%%%%%%%%%%%%%%%%%%%%%%%%%%%%%%%%%%%%%%%%%%%%
\subsection{Sturm-Liouville analysis}

Looking at the Fokker-Planck equation, Eq.~(\ref{warmFP}), we can
identify the following differential operator:

\begin{eqnarray} \label{FPoperator}
L_{FP} &=&    -\frac{\partial }{\partial \varphi}  D^{(1)}(\varphi)  -
D^{(1)}(\varphi) \frac{\partial }{\partial \varphi}  \nonumber \\ &+&
\frac{\partial^2 }{\partial \varphi^2} D^{(2)}  +  \frac{\partial
}{\partial \varphi} D^{(2)}  \frac{\partial }{\partial \varphi}
+D^{(2)}  \frac{\partial^2 }{\partial \varphi^2}\;,
\end{eqnarray}
which, in the light of Eq.~(\ref{warmFP}), allow us to write the
differential equation

\begin{equation} \label{current}
L_{FP} P(\varphi,t) = -  \frac{\partial }{\partial \varphi}
S(\varphi,t) \;,
\end{equation}
where $S(\varphi,t)=D^{(1)} P(\varphi,t)  -  \frac{\partial }{\partial
  \varphi} \left[D^{(2)}  P(\varphi,t) \right]$ is called the
probability current. 

We can write the general solution of the {}Fokker-Planck equation,
Eq.~(\ref{current}), as

\begin{equation} \label{solgeral}
P(\varphi,t) = \sum_{n} C_{n} P_n(\varphi) e^{\Lambda_n t}\;, 
\end{equation}
where $C_n$ are constants and the sum is performed over all
eigenvalues $\Lambda_n$ of the operator given by
Eq.~(\ref{FPoperator}), which in turn satisfies the following
eigenvalue equation:

\begin{equation} \label{TI}
L_{FP} P_n(\varphi) = \Lambda_n P_n(\varphi)\;.
\end{equation}

It is easy to show that the operator  $L_{FP}$,
Eq.~(\ref{FPoperator}), is not Hermitian. By redefining  variables
such that~\cite{Winitzki1996,winitzki2009eternal},

\begin{align} \label{rescaling}
& \varphi \rightarrow \int d \sigma\sqrt{D^{(2)}(\sigma)} \;,
  \nonumber \\ & \frac{\partial}{\partial \varphi} \rightarrow
  \frac{1}{\sqrt{D^{(2)}(\varphi)}} \frac{\partial}{\partial
    \sigma}\;, \nonumber \\ & P_n(\varphi) \rightarrow
  \frac{1}{D^{(2)}(\sigma)^{3/4}} \exp{\left[ \frac{1}{2} \int
      d \sigma \frac{D^{(1)}(\sigma)}{\sqrt{D^{(2)}(\sigma)}}    \right]}
  \psi_n(\sigma) \;,
\end{align}
we can transform the original {}Fokker-Planck equation into a
Sturm-Liouville problem. The advantage of this transformation rests in
the fact that the Sturm-Liouville operator is self-adjoint on the
Hilbert space, and its eigenvalues $\lambda$, $L_{SL}y_{\lambda}(x)=
\lambda y_{\lambda}(x)$, are real.  Inserting the above
transformations in Eq.~(\ref{TI}), we obtain a new eigenvalue
equation, which can be expressed as

\begin{equation} \label{schrodinger}
\frac{\partial^2}{\partial \sigma^2}\psi_n(\sigma) - V_S(\sigma)
\psi_n(\sigma) = \Lambda_n \psi_n(\sigma) \;, 
\end{equation}
where the effective potential $V_S$ is defined in terms of the drift
$D^{(1)}(\sigma)$ and diffusion $D^{(2)}(\sigma)$ coefficients.
For our purpose, we write $V_S$ in terms of the old variable $\varphi$, 
which gives

\begin{equation} \label{potentialschrodinger}
V_S(\varphi) = \frac{3}{16} \frac{(D^{(2)}_{,\varphi})^2}{D^{(2)}} -
\frac{D^{(2)}_{,\varphi \varphi}}{4} -
\frac{D^{(2)}_{,\varphi}D^{(1)}}{2 D^{(2)}} +
\frac{D^{(1)}_{,\varphi}}{2} +  \frac{(D^{(1)})^2}{4 D^{(2)}} \;.
\end{equation}

One can interpret Eq.~(\ref{schrodinger}) formally as a time
independent Schr\"odinger equation  describing a particle in a
potential $V_S$ with energy values $-\Lambda_n$. The same procedure can be
performed for the  volume weighted distribution $P_V(\varphi,t)$
equation, which is given by 

\begin{equation} \label{TIfisica}
\left[ L_{FP}+3H\right] P_{V_n}(\varphi) = \Lambda'_n
P_{V_n}(\varphi)\;,
\end{equation}
which adds a $-3H$ term to the effective potential,
Eq.~(\ref{potentialschrodinger}).

To analyze the {}Fokker-Planck equation,  one can make use of the
Sturm-Liouville theory \cite{Winitzki1996,winitzki2009eternal}.  The
Schr\"odinger-like equation for the comoving probability distribution
$P(\phi,t)$, Eq.~(\ref{schrodinger}), is a particular case of the
general Sturm-Liouville problem. Instead of considering $P(\phi,t)$
for our analysis, it is more useful to consider the volume weighted
distribution $P_V(\phi,t)$  due to its physical relevance.  {}For each
of these distributions, we can write the solution $\Psi(x,t) = \sum_n
\psi_n(x) e^{\Lambda_n t}$, with energy values $E_n = - \Lambda_n$. {}For the
distribution $P_V(\phi,t)$, we can write $P_V(\phi,t) = \sum_n
P_{V_n}(x) e^{\Lambda'_{n} t}$. If $\Lambda'_0>0$ ($E'_0<0$), the
physical volume of the inflating regions grows with time, and eternal
self-reproduction is present. Taking the boundary conditions into account, 
the following expression can be written for the zeroth eigenvalue 
\cite{Winitzki1996,winitzki2009eternal}:
\begin{equation} \label{Lambdamin}
\Lambda_0 =  - \mbox{min}_{\psi(\sigma)}  \frac{\int d\sigma \left[
    \left(\frac{d\psi_n}{d\sigma}\right)^2 + V_S(\sigma)
    \psi_n^2\right]}{\int d\sigma\; \psi_n^2}\,.
\end{equation}

{}From Eq.~(\ref{Lambdamin}), we can see that the only possibility
compatible with eternal self-reproduction is if there is at least a
range $\sigma_1<\sigma < \sigma_2$, such that the effective potential
$V_S(\sigma)$ is negative. Since the magnitude of the derivative term
in the numerator of Eq.~(\ref{Lambdamin}) is not determined, the
SLA of the Fokker-Planck equation cannot ensure the presence of eternal
inflation.  However, when we analyze $V_S(\sigma)$ together with the
FDR condition, Eq.~(\ref{condition}), a conclusive SLA can be performed.  

{}For the numerical analysis performed in Sec.~\ref{results}, we found
that it is more convenient to  analyze $V_S(\sigma)$ in terms of the
inflaton amplitude $\varphi$, instead of $\sigma$.   Since
inflationary dynamics is given in the variable $\varphi$, we use the
functional relation between $\varphi$ and $\sigma$, given by the first
expression in Eq.~(\ref{rescaling}), to write $V_S$ as a function of
$\varphi$.

%%%%%%%%%%%%%%%%%%%%%%%%%%%%%%%%%%%%%%%%%%%%%%%%%%%%%%%%%%%%%%%%%%%%%%%%%%%
\subsection{Eternal points analysis}

{}Finally, a third tool typically used to study the presence of eternal
inflation is to look for the presence of eternal points. Eternal
points are comoving worldlines {\bf x} that never reach the reheating
surface, i.e, are those points for which inflation ends at
$t=\infty$. Therefore, if one is able to proof the existence of
eternal points, eternal inflation occurs. The existence of eternal
points can be addressed  by solving a nonlinear diffusion  equation
for the complementary probability of having eternal points
\cite{winitzki2009eternal}

\begin{equation} \label{eternal_points}
D^{(2)}(\varphi)\overline{X}''(\varphi) +
D^{(1)}(\varphi)\overline{X}'(\varphi) + 3H(\varphi)
\overline{X}(\varphi) \ln{\overline{X}(\varphi)}=0 \; ,
\end{equation}
where prime indicates a derivative with respect to $\varphi$: $' \equiv
d/d\varphi$.  $\overline{X}$ is related to $X$ by $X =  1-
\overline{X}$, where $X$ is the probability of having eternal
points. Eternal points exist when there is a nontrivial solution for
$X(\varphi)$.  An approximate solution for Eq.~(\ref{eternal_points})
is obtained in the FDR neglecting the $D^{(1)}\overline{X}'$ term
when using the ansatz

\begin{equation} \label{ansatz}
\overline{X}(\varphi) = e^{-W(\varphi)} \;,
\end{equation}
where we have assumed $W$ to be a small varying function,
$W''\ll(W')^2$. In terms of $W(\varphi)$, Eq.~(\ref{eternal_points})
takes the form

\begin{equation} \label{ansatz_eq}
D^{(2)}(W')^2 - 3HW =0 \;.
\end{equation}
The solution to the above equation can be formally expressed as

\begin{equation} \label{ansatz_sol}
W(\varphi)=\frac{1}{4}\left(\ \int\limits_{\varphi_{\rm{th}}}^\varphi
\sqrt{ \frac{3H}{D^{(2)}} }d\varphi\right)^2 \;, 
\end{equation}
where $\varphi_{\rm{th}}$ is the threshold amplitude value for the
inflaton field, obtained from  Eq.~(\ref{condition}) and represents
the boundary of the fluctuation-dominated  range of $\varphi$ where
eternal inflation ends.

%%%%%%%%%%%%%%%%%%%%%%%%%%%%%%%%%%%%%%%%%%%%%%%%%%%%%%%%%%%%%%%%%%%%%%%%%%%%%%%%
\section{Generalizing eternal inflation in the context of warm inflation
dynamics}
\label{sec3}

In the first-principles approach to warm inflation, we start by
integrating over field degrees of freedom other than the inflaton
field.  The resulting effective equation for the inflaton field turns
out to be a Langevin-like equation with dissipative and stochastic
noise terms. An archetypal equation of motion for the inflaton field
can be written as \cite{WIreviews1,Berera:2007qm}

\begin{equation} \label{eomWI}
\ddot{\Phi}({\bf x},t) + \left[3H + \Upsilon \right] \dot{\Phi}({\bf
  x},t) - \frac{1}{a^2} \nabla^2 \Phi({\bf x},t) + V_{,\Phi}(\Phi) =
\xi_T({\bf x},t)\;, 
\end{equation}
where $\Upsilon = \Upsilon(\Phi,T)$ is the dissipation coefficient,
whose functional form depends on the specifics of the microphysical
approach (see, e.g., Refs.~\cite{Bastero-Gil2011,BasteroGil:2012cm}
for details), and $\xi_T({\bf x},t)$ is a thermal noise term coming
from the  explicit derivation of Eq.~(\ref{eomWI}) and that satisfies
the fluctuation-dissipation relation,

\begin{equation}
\langle\xi_T({\bf x},t)\xi_T({\bf x}',t')\rangle=  2 \Upsilon T
a^{-3}\delta^{(3)}({\bf x}-{\bf x}')\delta(t-t').
\label{snoise}
\end{equation}

{}Following the Starobinsky stochastic program, we perform a
coarse graining  of the quantum inflaton field $\Phi$, by decomposing
it into short  and long wavelength parts, $\Phi_{<}$ and $\Phi_{>}$,
respectively,

\begin{equation}
\Phi({\bf x},t)=\Phi_{<}({\bf x},t) + \Phi_{>}({\bf x},t).
\label{field_decomposition}
\end{equation}
In order to define $\Phi_{<}$, a filter function $W(k,t)$ is
introduced such that it eliminates the long wavelength modes ($k<a
H$), resulting in

\begin{equation} \label{modeexpansion}
\Phi_<({\bf x},t) \equiv \phi_q({\bf x}, t) = \int \frac{d^3 k}{(2
  \pi)^{3/2}} W(k,t)\left[\phi_{\bf k}(t) e^{-i{\bf k} \cdot {\bf x}}
  \hat{a}_{\bf k} + \phi_{\bf k}^{*}(t) e^{i{\bf k} \cdot {\bf x}}
  \hat{a}^{\dagger}_{\bf k}\right] \;,
\end{equation}
where  $\phi_{\bf k}(t)$ are the field modes in momentum space, and
$\hat{a}^{\dagger}_{\bf k}$ and $\hat{a}_{\bf k}$ are the creation and
annihilation operators,  respectively.  The simplest filter function
that is usually assumed in the literature has the form of a Heaviside
function,  $W(k,t)=\Theta(k-\epsilon a(t)H)$, where $\epsilon$ is a
small number.

Using the field decomposition defined in
Eq.~(\ref{field_decomposition}), we obtain the following equation for
the long wavelength modes:

\begin{equation} \label{eom_phi_long}
\ddot{\Phi}_{>}({\bf x},t) + 3H\left(1 + Q \right) \dot{\Phi}_{>}({\bf
  x},t) - \frac{1}{a^2} \nabla^2 \Phi_{>}({\bf x},t) +
V_{,\phi}(\Phi_{>}) = \xi_q({\bf x},t) + \xi_T({\bf x},t)\;, 
\end{equation}
where the quantum noise is given by

\begin{equation} \label{noise_qu}
\xi_q({\bf x},t) = - \left[\frac{\partial^2}{\partial t^2} + 3H\left(1
  + Q \right)\frac{\partial}{\partial t} - \frac{1}{a^2} \nabla^2 +
  V_{,\phi\phi}(\Phi_{>}) \right]\Phi_{<}({\bf x},t), \;
\end{equation}
and $Q=\Upsilon/3H$ is the dissipative ratio. The two-point
correlation function satisfied by the quantum noise  is given in the
Appendix~\ref{D2diss}. 

We are interested in a field that is nearly homogeneous inside a
H region, then we can consider the approximation $\Phi_{>}({\bf
  x},t)\approx\varphi(t)$.    In addition, we consider the slow-roll
approximation to obtain a Langevin-like equation of the form of
Eq.~(\ref{langevin}). {}From these requirements,
Eq~(\ref{eom_phi_long}) gives

\begin{equation} \label{eom}
\dot \varphi = f(\varphi)  + \sqrt{2D^{(2)}_{\rm{(vac)}}}\;
\zeta_q(t) + \sqrt{2D^{(2)}_{\rm{(diss)}}}\;  \zeta_T(t)  \;,
\end{equation}
where $\langle  \zeta_i(t) \zeta_i(t') \rangle = \delta(t-t')$ for
$i=q,T$, for the quantum and dissipative noises, respectively.  In
Eq.~(\ref{eom}),  the drift term is given by

\begin{equation}
\label{coeffs_WI_1}
f(\varphi) = -\frac{V_{,\varphi}(\varphi)}{3H (1 + Q)},
\end{equation}
while the diffusion coefficients are given by (see Appendix~\ref{D2diss},
for details)

\begin{align}
\label{coeffs_WI_2}
D^{(2)}_{\rm{(vac)}}  & = \frac{H^3}{8\pi^2} \left(1 + 2 n_{\tilde
  k}\right), \\   D^{(2)}_{\rm{(diss)}} & = \frac{H^2
  T}{80\pi}\frac{Q}{(1+Q)^2} ,
\end{align}
where $n_{\tilde k}$ is the statistical occupation number for the
inflaton field when in a thermal bath~\cite{LAS2013}, which is
evaluated at the lower limit scale separating the quantum and thermal
fluctuations, chosen as ${\tilde k}/a \approx T_H$, where $T_H= H/(2
\pi)$ is the Gibbons-Hawking temperature.

Equation~(\ref{eom}) is the warm inflationary analogous to the
Starobinsky  cold inflation one, Eq.~(\ref{eomEI}), now accounting for
the backreaction of both quantum and thermal noises (see, e.g.,
Ref.~\cite{LAS2013}, where these equations are explicitly derived in
the context of warm inflation for more details).

Equation (\ref{eom}) reduces to Eq.~(\ref{eomEI}) in the cold
inflation limit, where $Q \to 0,  T\to 0, n_{\tilde k} \to 0$. Some
other useful limiting cases of Eq.~(\ref{eom}) are: i) the
\textit{weak warm inflation} (WWI) limit $Q\ll 1,\ T/H\ll1$; ii) the
\textit{weak dissipative warm inflation} (WDWI) limit $Q\ll
1,\ T/H\gg1$; and   iii) the \textit{strong dissipative warm
  inflation} (SDWI) limit $Q\gg 1,\ T/H\gg1$. {}For example, writing
$n_{\tilde k} = 1/[\exp(T_H/T)-1]$, in the WDWI limit, we have that

\begin{equation} \label{eom_sQ_lT}
\dot \varphi \approx - \frac{V_{,\varphi}(\varphi)}{3H} +
\frac{H^{3/2}}{2\pi}\sqrt{2\frac{T }{T_H}}   \zeta_q(t)  \;,
\end{equation}
while in the SDWI limit, we obtain that

\begin{equation} \label{eom_lQ_lT}
\dot \varphi \approx - \frac{V_{,\varphi}(\varphi)}{3HQ} +
\frac{H^{3/2}}{2\pi}\sqrt{2\frac{T }{T_H}} \zeta_q(t)  \;.
\end{equation}

{}From Eqs.~(\ref{eom_sQ_lT}) and (\ref{eom_lQ_lT}), one notices that
the drift coefficient is attenuated due to the presence of
dissipation, whereas dissipation plays no significant role for
diffusion in both $Q\ll1$ and $Q\gg1$ limits. The opposite situation
happens when accounting for the effect of the temperature, which
always tends to enhance the diffusion  coefficient in warm inflation,
while its effects on the drift term is only manifest through the
dependence of the dissipation coefficient on the temperature.

The homogeneous background inflaton field is defined as the
coarse-grained field integrated in a H region volume:
$\phi(t)=(1/V_H)\int d^3{\bf x}\ \Phi({\bf x},t)$, where $V_H=\frac{4 \pi}{3H^3}$.
The background equation of motion for $\phi(t)$ becomes

\begin{equation} \label{eom_inflation_back}
\ddot{\phi}(t) + 3H(1+Q)\dot{\phi}(t) + V_{,\phi} = 0\;.
\end{equation}
The radiation energy density produced during warm inflation is
described by the evolution equation

\begin{equation} \label{eom_radiation_back}
\dot{\rho}_R + 4H\rho_R  = \Upsilon \dot{\phi}^2\;,
\end{equation}
where $\rho_R=C_R T^4$, $C_R=\pi^2g_*/30$ and $g_*$ is the effective
number of light degrees of freedom\footnote{In all of our numerical
  results, we will assume for $g_*$ the Minimal Supersymmetric Standard
  Model value  $g_*\approx228.75$ as a representative value. In any
  case, our results are only weakly dependent on the precise value of
  $g_*$.}. In the slow-roll regime, Eqs.~(\ref{eom_inflation_back})
and (\ref{eom_radiation_back}) can be approximated to

\begin{align} \label{eoms_sra}
 3H(1+Q)\dot{\phi} &\simeq - V_{,\phi} \;,  \\
 \label{eoms_sra2}
 \rho_R  &\simeq  \frac{3}{4}Q\dot{\phi}^2 \;,
\end{align}
while the slow-roll conditions in the warm inflation case are given by

\begin{align} 
\label{sraparameters}
\varepsilon &= \frac{1}{16\pi G} \left( \frac{V_{,\varphi}}{V}
\right)^2 < 1+Q \;, \\ \eta &= \frac{1}{8\pi G}  \frac{V_{,\varphi
    \varphi}}{V}  < 1+Q \;, \\ \beta & = \frac{1}{8\pi G}
\frac{\Upsilon_{,\varphi} V_{,\varphi}}{\Upsilon V}   < 1+Q \;,
\end{align}
where $G=1/(8\pi M_p^2$) is the Newtonian gravitational constant and
$M_p = m_{\rm{p}}/\sqrt{8\pi}$ is the reduced Planck mass.

In this work, we will be using in our analysis monomial forms for the
inflaton, which are the chaoticlike and hilltoplike potentials. The
chaoticlike potentials are defined as

\begin{equation} \label{potential}
V(\varphi) =  V_0\left( \frac{\varphi}{M_{p}}\right)^{2n} \;,
\end{equation}
where $n$ is a positive integer. The other class of potentials are the
hilltop ones~\cite{boubekeur2005hilltop}, with potential defined as

\begin{eqnarray}\label{potential_hybrid_nneq0}
V(\varphi) =
V_0\left[1-\frac{|\gamma|}{2n}\left(\frac{\varphi}{M_p}\right)^{2n}\right]\;.
\end{eqnarray}
In Eqs.~(\ref{potential}) and (\ref{potential_hybrid_nneq0}),
$V_0=\lambda M_{p}^4/(2n)$ and $\gamma$ is a free parameter. Here, we
will consider the cases  for $n=1$~(quadratic), $n=2$~(quartic), and
$n=3$~(sextic) chaotic potentials, whereas for the hilltop potential,
we will study  the cases for $n=1$~(quadratic) and $n=2$~(quartic),
for some values of the constant $\gamma$ motivated by the recent
Planck analysis for these type of potentials~\cite{Planck2015}.  Note
that the hilltop potential, Eq.~(\ref{potential_hybrid_nneq0}), is
usually written in the literature as $V=\Lambda^4
\left(1-\varphi^p/\mu^p\right)$. Thus, we identify $V_0=\Lambda^4$,
$p=2n$ and $\mu^{2n}=(2n/|\gamma|)M_p^{2n}$ for comparison.
Note that chaotic monomial potentials for the inflaton, in the cold inflation picture, 
are highly (for the quartic and sextic cases) or marginally (for the quadratic
case) disfavored by the Planck data. However, they are still in agreement with 
the Planck data in the context of warm inflation (see, e.g., Ref.~\cite{LAS2013} and, 
in particular, Ref.~\cite{bartrum2014importance} for a detailed analysis for
the case of the quartic chaotic potential 
in warm inflation). This is why we have included the potentials of the form of
Eq.~(\ref{potential}) in our analysis. 
On the other hand, hilltop potentials are found to be in agreement with the 
Planck data in both cold and warm inflation pictures. Both chaotic and hilltop
potentials are also representative examples of large field (chaotic)
and small field (hilltop) models of inflation. We thus expect that other
forms of potentials that fall into those categories should also have similar
results to the ones we have obtained using the above form of potentials.

{}For the dissipation coefficient $\Upsilon$ appearing in the inflaton
effective equation of motion, we will consider the microscopically
motivated form, that is a function of the temperature and the inflaton
amplitude, given
by~\cite{WIreviews1,Bastero-Gil2011,BasteroGil:2012cm}

\begin{equation}\label{upsilon}
\Upsilon = C_{\varphi} \frac{T^3}{\varphi^2}\;, 
\end{equation}
where $C_{\varphi}$ is a dimensionless dissipation parameter that
depends on the specifics of the interactions in warm inflation. The
dissipation coefficient Eq.~(\ref{upsilon}) is obtained in the
so-called low temperature regime for warm inflation. {}For example,
this form of dissipation can be derived for the case of a
supersymmetric model for the inflaton and the interactions, 
whose superpotential
is of the
form, $W = g \Phi X^2/2 + h X Y_i^2/2$, with chiral superfields
$\Phi$, $X$, and $Y_i$, $i = 1, ..., N_Y$. In the regime where the
$X$ fields have masses larger than the temperature and $Y_i$ are light
fields, $m_Y \ll T$, we have that~\cite{BasteroGil:2012cm}:
$C_{\varphi} \simeq 0.02 h^2 N_Y$.

It is worth to call attention to the fact that depending on the chosen
initial conditions for $\varphi$ and $Q$, inflation can begin in some
dissipative regime and end in another one. In chaotic inflation, $Q$ is
a quantity that {\it always} increases with time. If one starts at the
WWI or WDWI regimes, it is possible to occur a dynamical transition to
the SDWI regime as the dynamics proceeds. {}For example, if the system
starts in the WDWI regime, there are two possibilities: the system
remains in the WDWI regime until the end of inflation, or it enters in
the SDWI regime before its end. Therefore, if these dissipative
dynamical transitions occur, the only natural direction is
WWI$\rightarrow$WDWI$\rightarrow$ SDWI.  On the other hand, in the
case of hilltop inflation, it can happen that $Q$ decreases with
time. Thus, transitions between regimes can occur in the opposite
direction to that in the case of chaotic inflation:
SDWI$\rightarrow$WDWI$\rightarrow$ WWI.

In warm inflation, dissipation and temperature effects can enhance or
suppress eternal inflation depending on the regime we are
analyzing. On one hand, we expect that thermal fluctuations, similar
to the role played by quantum fluctuations in cold inflation, should
enhance eternal inflation. But dissipation can act in the opposite
direction, by damping the fluctuations and regulating the rate at
which energy from the inflaton field is transferred to the radiation
bath, acting as a suppressor of eternal inflation. In our numerical
results, we will see the nontrivial effects from these two opposite
quantities, which can be expressed in terms of the dissipation ratio
$Q= \Upsilon/(3 H)$ and the temperature ratio $T/H$.  {}For
convenience, these quantities are expressed in terms of their values
at a horizon crossing, since this is the point we can make contact with
observational constraints.  {}For instance, the primordial power
spectrum at a horizon crossing can be written
as~\cite{bartrum2014importance,LAS2013}

\begin{equation} \label{newP}
\Delta_{{\cal R}}^{(\rm{tot})} = \Delta^{(\rm{vac})}_{\rm{T}} +
\Delta^{(\rm{diss})} = \left(\frac{ H_{*}}{\dot{\phi}}\right)^2
\left(\frac{ H_{*}}{2\pi}\right)^2 \left[1 + 2 n_{*} +
  \left(\frac{T_{*}}{ H_{*}}\right)  \frac{2\sqrt{3}\pi Q_{*}}{\sqrt{3
      + 4 \pi Q_{*}}}\right] \;,
\end{equation}
where $\Delta^{(\rm{vac})}_{\rm{T}}=\Delta^{(\rm{vac})}\left(1 + 2
n_{*}\right)$ is the vacuum power spectrum of cold inflation
$\Delta^{(\rm{vac})}$ with the enhancement due to a nonvanishing
statistical distribution for the inflaton field in the thermal bath,
$n_{*} \equiv n_{k_{*}}$. The term $\Delta^{(\rm{diss})}$ is the
contribution to the power spectrum due to dissipation. All quantities
in Eq.~(\ref{newP}) are evaluated at the scale of a horizon crossing,
with $k_* = a_*H_*$.  We will assume that the distribution function
$n_{k_*}$ for the inflaton is that of thermal equilibrium and, thus,
is given by the Bose-Einstein distribution form, $n_{k_*} =
1/[\exp(H_*/T_*) -1]$. This assumption obviously depends on the
details of the microphysics involved during warm inflation. Some
physically well motivated interactions of the inflaton field with
other degrees of freedom  during warm inflation, that are able to
bring the inflaton to thermal equilibrium with the radiation bath,
have been discussed in
Refs.~\cite{BasteroGil:2012cm,bartrum2014importance}.  In this work we
will not consider further these possible details involving model
building in warm inflation, but we will consider both possibilities,
of an inflaton in thermal equilibrium, thus with a Bose-Einstein
distribution form, and also the case where the inflaton might not be
in thermal equilibrium with the radiation bath, in which case, it might
have a negligible statistical distribution $n_{k} \approx 0$.  Note
that in the limit $(Q_{*},T_{*},n_{*})\rightarrow 0$, one recovers the
standard cold inflation primordial spectrum as expected,
$\Delta_{{\cal R}}= H^4/(4 \pi^2 \dot\phi^2)$. 

Recently~\cite{bastero2014cosmological}, it was also shown that by
accounting for noise effects in the radiation bath in the perturbation
expressions, there can be an additional enhancement of the spectrum in
the dissipation term in Eq.~(\ref{newP}) by a factor of ${\cal
  O}(40)$, giving

\begin{equation} \label{newP_RADNOISE}
\Delta^{(\rm{diss})} \to \Delta^{(\rm{diss})}_{\rm{RN}}    \approx
\left(\frac{ H_{*}}{\dot{\phi}}\right)^2 \left(\frac{
  H_{*}}{2\pi}\right)^2  \frac{T_{*}}{ H_{*}} \frac{80\sqrt{3}\pi
  Q_{*}}{\sqrt{3 + 4 \pi Q_{*}}} \;.
\end{equation}
{}For the numerical analysis shown in the next section, we will
consider the power spectrum given by Eq.~(\ref{newP}), but we also
consider the correction (\ref{newP_RADNOISE}) due to the possibility
of extra random terms in the full perturbation equations.  This,
together with the considerations on $n_k$ explained above, will help
us to better assess the effects that these contributions have on the
emergence of eternal inflation in warm inflation.

The expression for the primordial spectrum given above,
Eq.~(\ref{newP}), or with the correction given by
Eq.~(\ref{newP_RADNOISE}) is a good fit for the complete numerical
result obtained from the complete set of perturbation equations in
warm inflation~\cite{bastero2014cosmological} for small values of $Q_*
\lesssim 0.1$. In our numerical studies, we will restrict the analysis
up to this value of dissipation ratio, though it could be extended to
larger values of $Q_*$ by coupling the equations to those of the full
perturbation equations, but we refrain to do this given the numerical
time consuming involved. Besides, the analysis for $Q_* \lesssim 0.1$
will already suffice to make conclusions on the nontrivial effects
that dissipation, noise, and the thermal radiation bath will have in
the emergence of a SRR in warm inflation.
 
Given the primordial spectrum, the model parameters, including those
for the inflaton potentials we consider in this work,
Eqs.~(\ref{potential})   and (\ref{potential_hybrid_nneq0}), are then
constrained such that they satisfy the amplitude of scalar
perturbations, $\Delta_{{\cal R}}\simeq 2.25 \times 10^{-9}$, in
accordance to the recent data from Planck~\cite{Planck2015}.

\begin{figure}[htb!]
\vspace{0.75cm} \centerline{
  \psfig{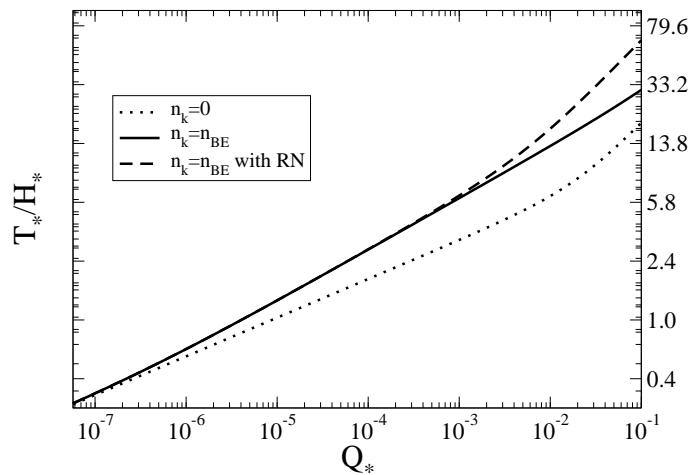} }
\caption{$T_{*}/H_{*}$ as a function of $Q_{*}$. The results are shown
  for two particular choices of the particle distribution, $n_k=0$ and
  for $n_k=n_{\rm{BE}}$, and by also accounting for the effects of the
  radiation noise correction in the power spectrum,
  Eq.~(\ref{newP_RADNOISE}) and in the absence of these effects,
  Eq.~(\ref{newP}). }
\label{vstar_Qstar}
\end{figure}

Note that from the evolution equations,
Eqs.~(\ref{eom_inflation_back}) and (\ref{eom_radiation_back}), with
$\Upsilon$ defined by Eq.~(\ref{upsilon}), and the constraint on the
inflaton potential given by the normalization on the amplitude of the
primordial spectrum,  one obtains a functional relation between $Q_*$
and $T_*/H_*$.  In {}Fig.~\ref{vstar_Qstar}, we plot the functional
relation between $T_{*}/H_{*}$ and $Q_{*}$. In this figure, we also
consider the cases where the particle distribution is given by $n_k=0$
and where radiation noise contribution to the power spectrum is taken
into account, for future reference. It is important to highlight that
the curves $T_{*}/H_{*}\times Q_{*}$ are approximately
potential independent  and they are also only mildly dependent on
$g_*$. Thus, {}Fig.~\ref{vstar_Qstar} also represents the functional
relation for the hilltop potentials used in the analysis done in the
next section.

Given the relation between $T_{*}/H_{*}$ and $Q_{*}$, we are free to
choose one of  these variables when presenting our analysis. 
We choose $Q_*$, since it is the most  transparent one, and for the 
corresponding values of $T_*/H_*$ for each value of $Q_*$ in the analysis, 
the reader is referred to consult the results of Fig.~\ref{vstar_Qstar}.
Thus, the effects of $Q$ and $T$ on the establishment of a SRR can be adequately 
addressed and contrasted with the cold inflation case. {}Further details about the
way we perform the numerical analysis are also explained in 
Appendix~\ref{numerical_analysis}.

In the particular case of Eq.~(\ref{eom}), we consider the thermal and
quantum noises as uncorrelated ones, since they have distinct origins
(see, e.g., Ref.~\cite{LAS2013}). This corresponds to assume that the
noises in Eq.~(\ref{langevin}) are uncorrelated, i.e, $\theta=0$.
Then, comparing the Langevin equations given by Eq.~(\ref{langevin})
with Eq.~(\ref{eom}) and the  coefficients given by
Eq.~(\ref{coeffs_FP}) with Eqs.~(\ref{coeffs_WI_1}) and
(\ref{coeffs_WI_2}), the drift and diffusion Fokker-Planck
coefficients are, respectively, given as follows:

\begin{align} 
& D^{(1)} =  -\frac{V_{,\varphi}(\varphi)}{3H (1 + Q)}
  \;, \label{arraste}  \\  & D^{(2)} =   \frac{ H^3}{8\pi^2} \left[1 +
    2 n_{\tilde k} + \left(\frac{T}{H}\right) \frac{\pi
      Q}{10(1+Q)^2}\right] \;.
\label{diffusion}
\end{align}
Then, starting from Eq.~(\ref{eom}), it is possible to derive a
{}Fokker-Planck equation that preserves the form of the original model
given by Eq.~(\ref{FPcold}). 

%%%%%%%%%%%%%%%%%%%%%%%%%%%%%%%%%%%%%%%%%%%%%%%%%%%%%%%%%%%%%%%%%%%%%%%%%%
\section{Results} \label{results}

To assess the effects of dissipation and thermal fluctuations on the
presence or absence of a SRR, we consider the tools described in the
previous section. Thus, we will be making use of the
effective potential $V_S$, the counting of H regions, the threshold
inflaton field $\phi_{\rm{th}}$, and the threshold number of {\it e}-folds
$N_{\rm{th}}$ in terms of the dissipation ratio $Q$ and $T/H$.  The
analysis of $V_S$ and of the counting of H regions produced in the SRR
are presented in parallel as complementary, as well 
as the analysis of $\varphi_{\rm{th}}$ and $N_{\rm{th}}$.

{}For the whole analysis, we have used the FDR condition,
Eq.~(\ref{condition}), to determine the regions of parameters for
which eternal inflation occurs.  This condition is our main tool of
analysis, which will become more transparent 
represented graphically by the aforementioned variables.

In warm inflation, the analysis of $V_S$ and the counting of H regions
are performed in the case where inflaton particles rapidly thermalize
and are given  by a Bose-Einstein distribution, $n_k=n_{\rm{BE}}$. The
analysis of $\varphi_{\rm{th}}$  and $N_{\rm{th}}$, additionally
consider the possibility where the inflaton particle distribution  is
negligible, $n_k=0$. It is also analyzed the case where we consider
the radiation noise (RN) contribution to the power spectrum,
represented by the enhancement given in Eq.~(\ref{newP_RADNOISE}).

It is useful to write Eq.~(\ref{eom}) and all related quantities in
terms of dimensionless variables. We introduce the following set of
transformations that we will be considering throughout this work: 

\begin{eqnarray}
&& \varphi = M_{p} x,  \;\;\;\; V = \lambda M_{p}^4 v/(2n), \;\;\;\; H
  = \lambda^{1/2} M_{p} L/(\sqrt{6n}), \nonumber \\  && T =
  \lambda^{1/2} M_{p} T'/(\sqrt{6n}), \;\;\;\; \Upsilon =
  \lambda^{1/2}M_{p}\Upsilon'/\sqrt{6n}, \nonumber \\ && \zeta_T =
  (6n)^{1/4} \lambda^{1/4} M_{p}^{1/2}  \zeta_T'/\sqrt{3}, \;\;\;\;
  \zeta_q = (6n)^{1/4} \lambda^{1/4} M_{p}^{1/2}  \zeta_q'/\sqrt{3},
  \nonumber \\ && t = 3t'/(\sqrt{6n}\lambda^{1/2}M_{p}).
\label{newvariables}
\end{eqnarray}
{}For example, in terms of the dimensionless variables defined above,
the dissipation coefficient $\Upsilon$, Eq.~(\ref{upsilon}) is written
as

\begin{equation}\label{gamma_dependent}
\Upsilon' = \frac{C_{\varphi}\lambda}{6n} \frac{T'^3}{x^2}\;. 
\end{equation}

The evolution of the inflaton field, Eq.~(\ref{eom}), expressed in terms of the
drift and diffusion coefficients, Eqs.~(\ref{arraste}) and
(\ref{diffusion}),  in terms of the
dimensionless variables (\ref{newvariables}) becomes 

\begin{equation} \label{EoM}
\frac{\partial x}{\partial t'}  =  - \frac{v_{,x}}{2nL(1+Q)}  +
\frac{\sqrt{3\lambda}}{6n} \frac{L^{3/2}}{2\pi}\left(1+2n_{\tilde k}
\right) \zeta_q' + \frac{\sqrt{3\lambda}}{6n}
\frac{L^{3/2}}{2\pi}\left(\frac{T'}{L}\right)\frac{\pi Q}{10(1+Q)^2}
\zeta_T'\;.
\end{equation}
{}From Eq.~(\ref{EoM}), we find that the volume weighted probability
distribution is the solution of the following dimensionless Fokker-Planck
equation: 

\begin{eqnarray} \label{FPdim}
\frac{\partial }{\partial t'} P_V(x ,t') &=&  \frac{\partial
}{\partial x} \left[\frac{v_{,x}}{2nL(1+Q)}  P_V(x,t') \right]
\nonumber\\ &+& \frac{\partial^2 }{\partial x^2}  \left\{
\frac{\lambda}{12n^2}\frac{L^{3}}{8\pi^2}\left[1+2n_{\tilde k}+
  \left(\frac{T'}{L}\right)\frac{\pi Q}{10(1+Q)^2}\right]  P_V(x,t')
\right\} \nonumber\\ &+&\frac{3L}{2n} P_V(x,t') \;. 
\end{eqnarray}

Using the dimensionless variables introduced in Eq.~(\ref{rescaling})
into Eq.~(\ref{FPdim}), it is possible to rewrite the  {}Fokker-Planck
equation (\ref{FPdim}) into a Schr\"odinger-like equation, whose
effective potential is given by

\begin{equation} \label{potentialschrodingeradm}
V_S(\sigma) = \frac{3}{16} \frac{(D^{(2)}_{,x})^2}{D^{(2)}} -
\frac{D^{(2)}_{,x x}}{4} - \frac{D^{(2)}_{,x}D^{(1)}}{2 D^{(2)}} +
\frac{D^{(1)}_{,x}}{2} +  \frac{(D^{(1)})^2}{4 D^{(2)}} -
\frac{3L}{2n}\;.
\end{equation}
In all
situations, we take into account the field backreaction on geometry,
since we are primarily interested in studying the global structure of
the inflationary universe. 

In the next subsections, we use the SLA to extract the relevant
information  from the above effective potential $V_S$, in the cold and
warm inflation cases,  and for both types of inflaton potentials
considered in this work, given by  Eqs.~(\ref{potential}) and
(\ref{potential_hybrid_nneq0}).  The analysis of $V_S$ is performed
comparatively with the number of H regions,
$\exp{(3)}\times(N_e-N_{\rm{th}})$, for a total number of {\it e}-folds
$N_e>N_{\rm{th}}$,  which gives the counting of H regions produced in
the FDR.

We will omit the analysis of $\bar{X}$ in the warm inflation case
because, as we will see in the following, this analysis is
qualitatively equivalent to the one provided by $V_S$, while in  cold
inflation, we present both for the sake of completeness.

In the following, we present results for the cold inflation scenario, 
for each inflaton potential model considered. Thereafter, we will
extend these results to include the effects of
dissipation and thermal radiation  in order to establish whether
they can enhance or suppress eternal inflation.

%%%%%%%%%%%%%%%%%%%%%%%%%%%%%%%%%%%%%%%%%%%%%%%%%%%%%%%%%%%%%%%%%%%%%%%%%
\subsection{Chaotic and hilltop models in the cold inflation case}
\label{cold_case}

As a warm up, let us apply the methods described in 
Sec.~\ref{sec2} to characterize eternal inflation for the case of
cold inflation, i.e., initially in the case of absence of thermal and
dissipative effects.  In the cold inflation case, the evolution of the
inflaton field, Eq.(\ref{EoM}) is given by 

\begin{equation} \label{cold_EoM}
\frac{\partial x}{\partial t'} = - x^{n-1} +
\frac{\sqrt{3\lambda}}{6n} \frac{x^{3n/2}}{2\pi} \zeta_q'  \;,
\end{equation}
where the dimensionless variables (\ref{newvariables}) were
used. Then, the volume weighted probability distribution  is the
solution of the following {}Fokker-Planck equation:

\begin{equation} \label{coldFPdim}
\frac{\partial }{\partial t'} P_V(x ,t') =   \frac{\partial }{\partial
  x} \left[x^{n-1}  P_V(x,t') \right]  + \frac{\partial^2 }{\partial
  x^2}  \left[ \frac{\lambda}{12n^2}\frac{x^{3n}}{8\pi^2}  P_V(x,t')
  \right] +\frac{3x^n}{2n} P_V(x,t') \;, 
\end{equation}
which is a particular case of Eq.~(\ref{FPdim}). Using the drift and
diffusion coefficients of Eq.~(\ref{coldFPdim})  in
Eq.~(\ref{potentialschrodingeradm}), the explicit form of the
effective potential $V_S$ is promptly obtained,  for both the chaotic
and the hilltop potentials.

{}For the chaotic model, the effective potentials for $n=1$, $n=2$, and
$n=3$ become, respectively,

\begin{align} \label{Veffcold}
& V_{S,\rm{chaotic}}^{n=1} =    - \frac{3}{2}x  + \frac{\lambda}{512
    \pi^2} x + \frac{3}{2}x^{-1}  + \frac{24\pi^2}{\lambda} x^{-3} \;,
  \nonumber \\ & V_{S,\rm{chaotic}}^{n=2} =   - \frac{\lambda}{512
    \pi^2} x^4  - \frac{3}{4}x^2  + \frac{5}{2}   + \frac{96
    \pi^2}{\lambda}x^{-4}  \;, \nonumber \\ & V_{S,\rm{chaotic}}^{n=3}
  =  - \frac{5 \lambda}{1536 \pi^2} x^7 - \frac{1}{2}x^3  +
  \frac{7}{2}x  + \frac{216\pi^2}{\lambda}x^{-5}  \;.
\end{align} 
{}For the hilltop model, we obtain that

\begin{align} \label{Veffcold_HT}
& V_{S,\rm{hillop}}^{n=1} =   -\frac{3\sqrt {v}}{2}
  +{\frac{\gamma}{4\sqrt{v}}} +{\frac {\lambda\gamma\sqrt{v}}{256 {\pi
      }^{2}}} + \left( {\frac {5 \lambda}{2048\pi^{2}\sqrt{v}}}
  +{\frac{1}{2{v}^{3/2}}} +{\frac {6\pi^{2}}{\lambda{v}^{5/2}}}
  \right)\gamma^{2} {x}^{2} \;, \nonumber \\ & V_{S,\rm{hillop}}^{n=2}
  =     -\frac{3\sqrt{v}}{4} +  \frac{3}{8}\left( {\frac {\lambda
      \sqrt {v}}{128{\pi }^{2}}} +{\frac { 1}{\sqrt {v}}}
  \right)\gamma {x}^{2} + \left( {\frac {5\lambda}{8192{\pi }^{2}
      \sqrt {v}}} +{\frac {1}{4{v}^{3/2 }}}+{\frac {6{\pi
      }^{2}}{\lambda{v}^{5/2}}} \right) {\gamma}^{2}{x}^{6} \;,
\end{align} 
where $v$ is typically $v\lesssim 1$ during the FDR. 

{}From Eqs.~(\ref{Veffcold}) and (\ref{Veffcold_HT}), due to the
typical smallness of $\lambda$ and $\gamma$, one observes that in both
effective potentials the negative terms are dominant for high (low)
values of $x$ in chaotic (hilltop) inflation.  Since high (low) $x_0$
are the typical initial values for chaotic  (hilltop) inflation, these
negative terms dominate for adequate suitable values  of $x_0$.  {}For
chaotic inflation, as inflation evolves from high $x=x_0$ values to
smaller $x$,  the positive terms of order $\mathcal{O}(\lambda^{-1})$
increase and tend to become more relevant, whereas for hilltop
inflation, the terms of order $\mathcal{O}(\gamma)$ and
$\mathcal{O}(\lambda^{-1}\gamma^2)$ tend to increase as inflation
evolves from small $x=x_0$ values  to higher $x$.  These positive
terms continuously increase the values of $V_S$ to less negative ones
during inflation, which proceeds until $V_S>0$ at the end of
inflation. {}From Eq.~(\ref{Lambdamin}), we have discussed that
eternal inflation is possible to occur if there is an interval of
$\varphi$ (i.e., $\sigma$) where $V_S<0$, which can be achieved for
these different forms of effective potentials. {}For inflation
beginning at an initial field configuration that respects the FDR
condition, Eq.~(\ref{condition}), we obtain a sufficiently negative
$V_S$ for eternal inflation to occur and, as the effective potential
becomes less negative, eternal inflation eventually ceases for some
less negative $V_S$, when the FDR condition is no longer satisfied,
i.e., $\left(d\psi_n/d\sigma\right)^2$ dominates over $V_S$ in
Eq.~(\ref{Lambdamin}). 

Together with the obtained effective potentials, we use the
dimensionless  version of the FDR condition, Eq.~(\ref{condition}), to
obtain $x_{\rm{th}}$,  which is the (threshold) value of $x$ for which
the FDR ends. If the condition Eq.~(\ref{condition}) gives a
$x_{\rm{th}}$ between $x_{\rm{0}}$ and $x_{\rm{f}}$,  it means that a
FDR is present. In addition, for the value of $x_{\rm{th}}$ for  each
potential, we can obtain the respective threshold number of {\it e}-folds
$N_{\rm{th}}$.

%%%%%%%%%%%%%%%%%%%%%%%%%%  FIGS %%%%%%%%%%%%%%%%%%%%%%%%%%%%%%%%%%%%%%%%%%

\begin{figure}[hbt]
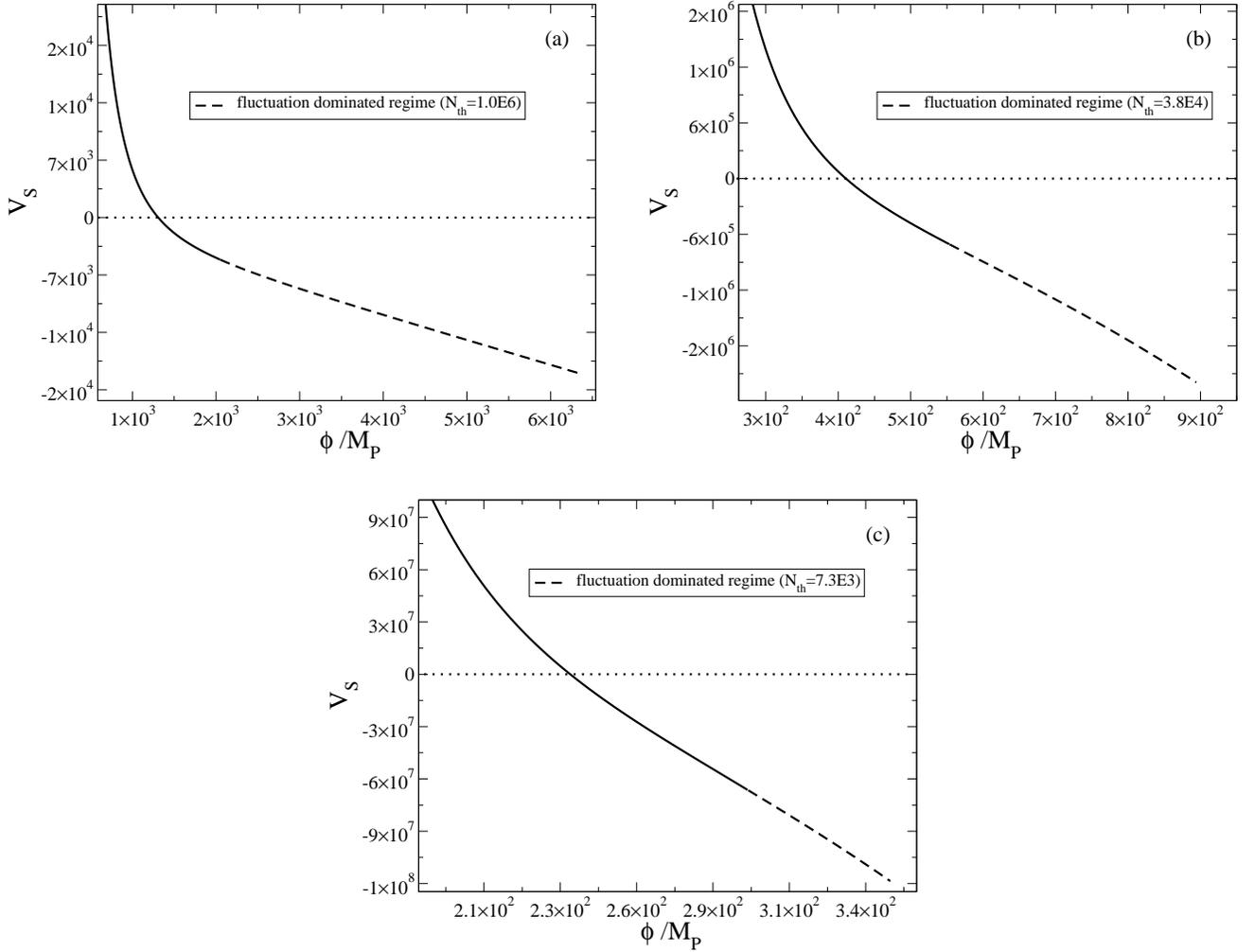

\vspace{0.75cm} \centerline{
  \psfig{file=fig2a.eps,scale=0.3,angle=0}
\hspace{0.5cm} \psfig{file=fig2b.eps,scale=0.3,angle=0} }
\vspace{0.5cm} \centerline{
  \psfig{file=fig2c.eps,scale=0.3,angle=0} }
%\vspace{0.25 cm}
\caption{The effective potential $V_S$ for: (a) quadratic, (b) quartic,
  and (c) sextic  chaotic inflation, respectively.  The dashed curves
  show the fluctuation-dominated range.  The values of $N_e$ chosen
  for each panel are given, respectively, by (a) $10^7$,  (b) $10^5$,
  and (c) $10^4$.}
\label{Veff_FDR_CI}
\end{figure}

\begin{figure}[htb]
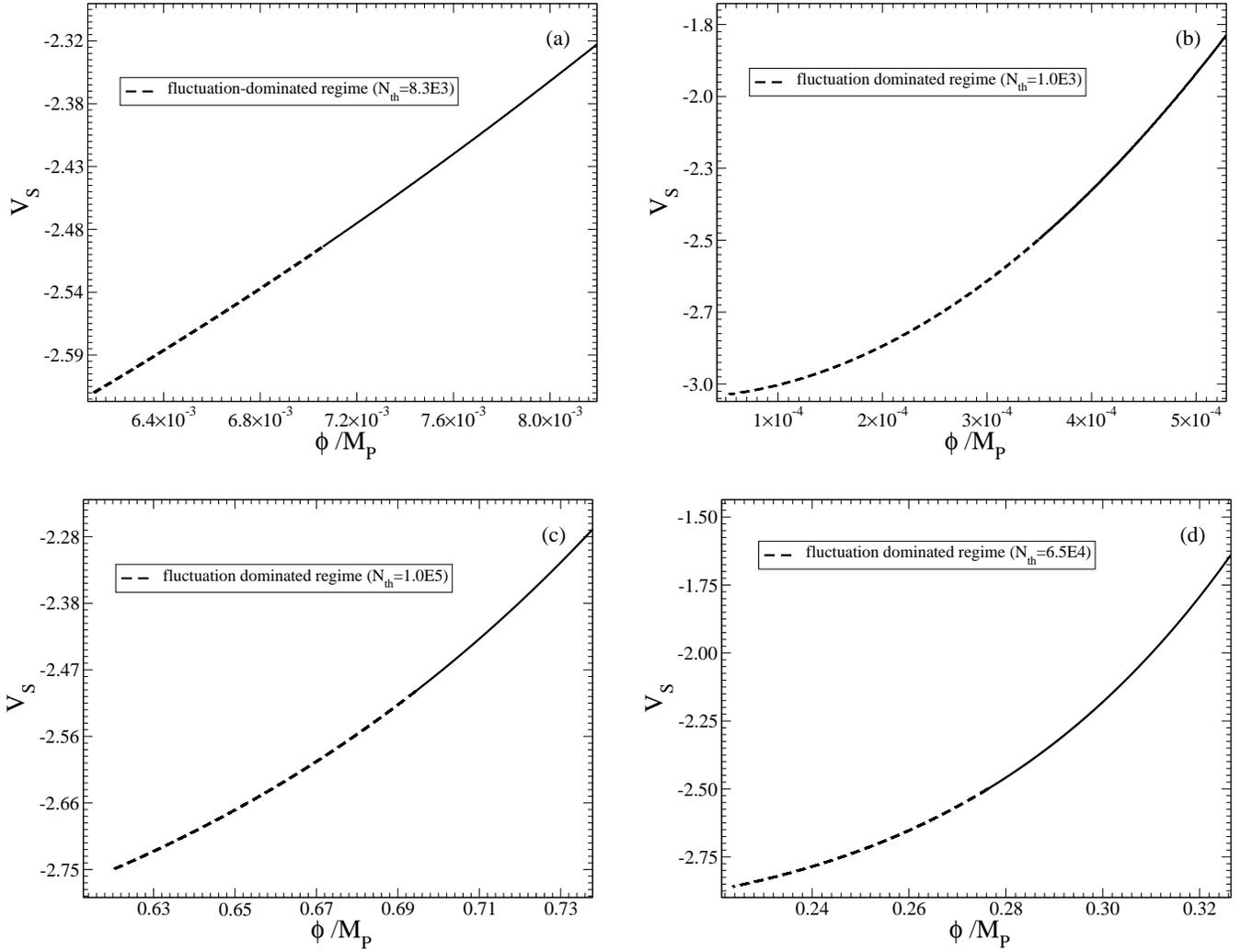

 \centerline{
   \psfig{file=fig3a.eps,scale=0.31,angle=0}
\hspace{0.5cm}
\psfig{file=fig3b.eps,scale=0.31,angle=0} }
\vspace{0.5cm} \centerline{
  \psfig{file=fig3c.eps,scale=0.31,angle=0}
\hspace{0.5cm}
\psfig{file=fig3d.eps,scale=0.31,angle=0} }
%\vspace{0.25 cm}
\caption{The effective potential $V_S$ for quadratic [panels (a) and
  (b)] and quartic  [panels (c) and (d)] hilltop inflation for some
  representative values of $\gamma$.  The dashed curves show the
  fluctuation-dominated range. The chosen values of $\gamma$ and
  $N_e$ for each panel are given, respectively,  by (a) $10^{-3}$ and
  $8.4\times10^3$, (b) $10^{-2}$ and $1.2\times10^3$, (c) $10^{-5}$
  and $1.3\times10^5$, and (d) $10^{-4}$ and $10^5$.}
\label{Veff_FDR_CI_HT}
\end{figure}

In  {}Figs.~\ref{Veff_FDR_CI} and ~\ref{Veff_FDR_CI_HT}, we show the
behavior of the effective potential $V_S$ for the chaotic and hilltop
inflation cases, respectively. Each curve represents an inflationary
evolution where we choose some $N_e>N_{\rm{th}}$, which means that
eternal inflation occurs, and is separated in dashed and solid lines
segments, which represent two distinct regimes. The dashed segment of
the negative part of $V_S$ corresponds to the FDR, which begins at the
lowermost points (the beginning of inflation and SRR, at $x=x_0$) and
ends at where dashed and solid line segments encounter (the end of
FDR, at $x=x_{\rm{th}}$).  The remaining part of the curves correspond
to the deterministic regime, which begins at $x=x_{\rm{th}}$ for
$V_S<0$ and ends in the topmost point at $x=x_{\rm{f}}$, where
inflation ends and $V_S>0$.  Particularly, in {}Fig~\ref{Veff_FDR_CI},
the initial point $[x_0,V_S(x_0)]$ is always the rightmost point on
the curve (recalling that $x$ decreases during the chaotic evolution),
and in {}Fig~\ref{Veff_FDR_CI_HT}, it is the leftmost point (recalling
that $x$ increases during the hilltop evolution). In both cases, the
vertical axis is constrained for a matter of scale, thus omitting the
final value $[x_f,V_S(x_f)]$. 

The dashed curves in {}Figs.~\ref{Veff_FDR_CI} and
\ref{Veff_FDR_CI_HT} represent the $N_e$-$N_{\rm{th}}$ {\it e}-folds of
eternal inflation where a SRR occurs, which means that for eternal
inflation to happen, inflation needs to begin at an initial inflaton
field value  adequate to provide the sufficient number of {\it e}-folds
$N_e>N_{\rm{th}}$.  The greater the length of the dashed curves, the
greater is the difference $N_e$-$N_{\rm{th}}$, indicating a stronger
SRR. In the opposite case, the smaller we choose $N_e$-$N_{\rm{th}}$ the
dashed line becomes smaller till it disappears for $N_e\leq
N_{\rm{th}}$, remaining the solid curve. {}For eternal inflation to
occur for the case of the chaotic inflation, the initial value for the
inflaton field, $\varphi_0$, needs to be  sufficiently large
($\varphi_0\gg M_P$), whereas for hilltop inflation it needs to be
sufficiently small ($\varphi_0\ll M_P$), i.e., very close to the top
of the potential at the origin.

The values of $x_{\rm{th}}$ given by the FDR condition (related to
each $N_{\rm{th}}$ shown in the figures) are $x_{\rm{th}}=2.0\times
10^3,\; 5.5\times10^2,\; 3.0\times 10^2$  in the chaotic cases
(Fig.~\ref{Veff_FDR_CI}) for  $n=1,2,3$, respectively, and in the
hilltop cases (Fig.~\ref{Veff_FDR_CI_HT}) by
$x_{\rm{th}}=7.0\times10^{-3}$ ($\gamma=10^{-3}$) and
$x_{\rm{th}}=3.5\times10^{-4}$ ($\gamma=10^{-2}$) for $n=1$, and
$x_{\rm{th}}=0.698$ ($\gamma=10^{-5}$) and $x_{\rm{th}}=0.276$
($\gamma=10^{-4}$) for $n=2$. In each case, we have set a value of
$N_e$ such that a SRR is viable, i.e., $V_S$ exhibits a negative
interval that contains a FDR.

%%%%%%%%%%%%%%%%%%%%%%%%%%%%%   FIG  %%%%%%%%%%%%%%%%%%%%%%%%%%%%%%%%%%%%%%%%

\begin{figure}[htb]
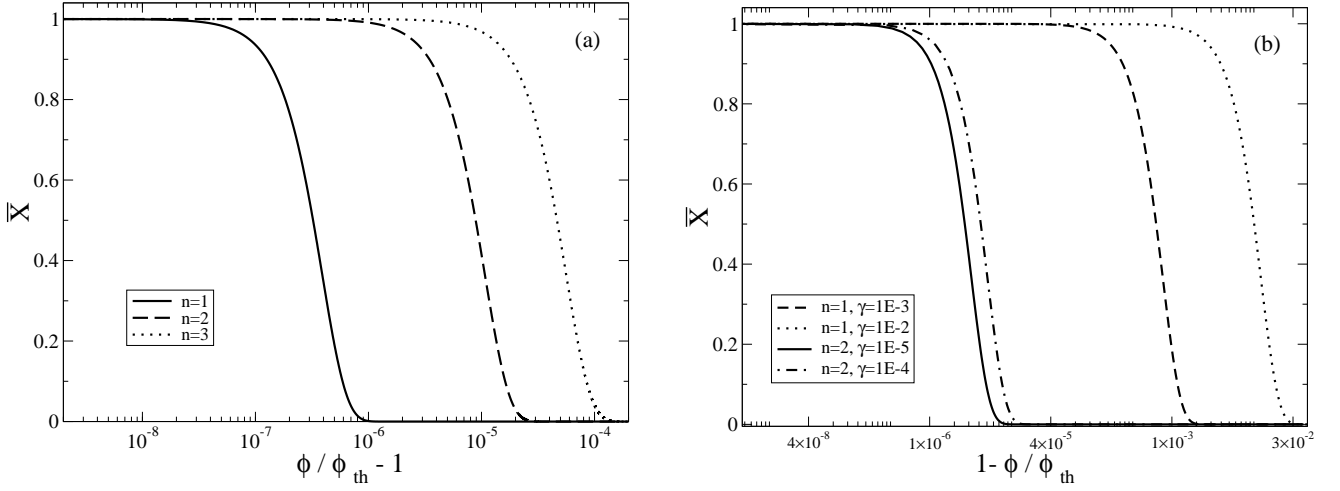

\vspace{0.75cm} \centerline{
  \psfig{file=fig4a.eps,scale=0.3,angle=0}
\hspace{0.5cm} \psfig{file=fig4b.eps,scale=0.3,angle=0} }
\caption{The probability of finding no eternal points in the chaotic
  [panel (a))  and hilltop (panel (b)] cold inflation cases.  The
  chosen values of $N_e$ are the same ones of the previous figures for
  each inflation case ($n=1,2,3$) and $\gamma$.}
\label{Xb_CI}
\end{figure}

In  the {}Fig.~\ref{Xb_CI}, we show the behavior of the probability of
having no eternal points for the chaotic [panel (a)] and hilltop
[panel (b)] potentials. In the horizontal axis, we plot values of $x$
between $x_{\rm{0}}$ and $x_{\rm{th}}$ (FDR range), suitably
parametrized by the variable $x/x_{\rm{th}}-1$ in the chaotic case
and $1-x/x_{\rm{th}}$, in the hilltop case. The initial points are at
the rightmost ones of the curves, indicating that eternal points are
initially present ($\bar{X}\approx 0$) and vanish at the end of FDR
($\bar{X}\approx 1$).

As expected, we observe that both tools we have used to assess the
presence of a SRR  produce results that are compatible between
them. Due to this compatibility, in the following  analysis, extended
to the case of warm inflation, we omit the study of eternal points for
the reason of  not being repetitive in performing both qualitative
$V_S$ and eternal points  analysis. In its place, we introduce the
counting of H regions versus dissipation  in parallel to the analysis
for $V_S$, which is a quantitative tool and more adequate for
describing the emergence of eternal inflation when considering  now
the effects of dissipation and radiation.

%%%%%%%%%%%%%%%%%%%%%%%%%%%%%%%%%%%%%%%%%%%%%%%%%%%%%%%%%%%%%%%%%%%%%%%%%%%%%
\subsection{Chaotic warm inflation}

Let us initially study the case of warm inflation with the chaotic
type of potentials.  {}For the polynomial potential in the warm
inflation case, the evolution of the inflaton field, Eq.~(\ref{EoM}),
is given by 

\begin{equation}
\frac{\partial x}{\partial t'}  =  - \frac{x^{n-1}}{1+Q}  +
\frac{\sqrt{3\lambda}}{6n} \frac{x^{3n/2}}{2\pi}\left(1+2n_{\tilde
  k}\right) \zeta_q'  + \frac{\sqrt{3\lambda}}{6n}
\frac{x^{3n/2}}{2\pi}\left(\frac{T'}{x^n}\right)\frac{\pi
  Q}{10(1+Q)^2} \zeta_T'\;.
\end{equation}
The volume weighted probability distribution is the solution of the
{}Fokker-Planck equation, Eq.~(\ref{FPdim}), given by 

\begin{eqnarray} 
\frac{\partial }{\partial t'} P_V(x ,t') &=&  \frac{\partial
}{\partial x} \left[\frac{x^{n-1}}{\left(1+Q\right)}  P_V(x,t')
  \right] \nonumber\\ &+& \frac{\partial^2 }{\partial x^2}  \left\{
\frac{\lambda}{12n^2}\frac{x^{3n}}{8\pi^2}\left[1+2n_{\tilde k} +
  \left(\frac{T'}{x^n}\right)\frac{\pi Q}{10(1+Q)^2}\right]  P_V(x,t')
\right\} \nonumber\\ &+&\frac{3x^n}{2n} P_V(x,t') \;. 
\end{eqnarray}
{}From the {}Fokker-Planck equation, we can obtain the effective
potential using Eq.~(\ref{potentialschrodinger}). The expression for
$V_S$ is too large to be presented in the text; thus, we chose to
present  some suitable representative results by numerical
integration.

%%%%%%%%%%%%%%%%%%%%%%%%%%%  FIG  %%%%%%%%%%%%%%%%%%%%%%%%%%%%%%%%%%%%%%%%%

\begin{figure}[htb!]
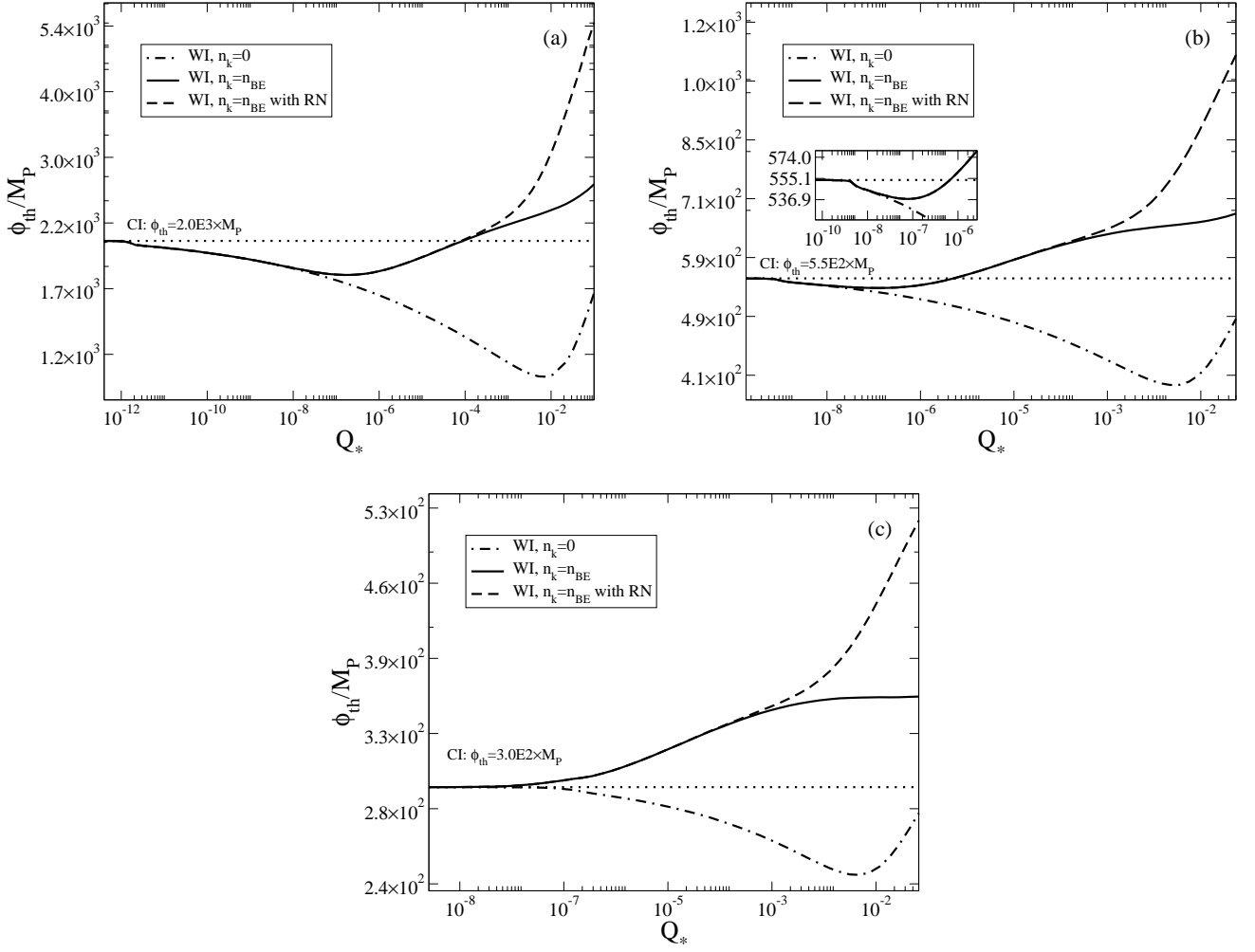

\vspace{0.75cm} \centerline{
  \psfig{file=fig5a.eps,scale=0.30,angle=0}
\hspace{0.5cm}
\psfig{file=fig5b.eps,scale=0.30,angle=0} }
\vspace{0.5cm}  \centerline{
  \psfig{file=fig5c.eps,scale=0.30,angle=0}}
\caption{ The threshold values of $\phi_{\rm{th}}$ versus $Q_{*}$ for
  the: (a) quadratic, (b) quartic, and (c) sextic chaotic inflation
  cases, respectively.  The solid (dash-dot) curves correspond to
  thermal (negligible) inflaton distribution $n_k$ while the dashed
  curve corresponds to the thermal inflaton distribution and also by
  accounting for radiation noise effects, given according to
  Eq.~(\ref{newP_RADNOISE}).}
\label{phith_Qstar}
\end{figure}

In {}Fig.~\ref{phith_Qstar}, we present the functional relation between
the threshold values of the inflaton field $\varphi_{\rm{th}}$ and
dissipation ratio $Q_*$.  As expected, in all panels, we recover the
cold inflation values  for sufficiently small $Q_{*}$.  The solid
curves in the panels represent the cases where the inflaton particle
distribution is given by the Bose-Einstein one. The additional
dash-dotted and  dashed curves represent the cases for which the
inflaton particle distribution is negligible and for which the
radiation noise contribution to the power spectrum is taken into
account, respectively.

{}For the quadratic potential ($n=1$), shown in panel (a) of
{}Fig.~\ref{phith_Qstar}, as we increase the value of $Q_{*}$, a
notable  nonlinear behavior emerges: for $Q_{*}$ approximately between
$10^{-12}$ and $7\times 10^{-5}$,  the condition for the presence of a
FDR is alleviated, since the threshold value $\varphi_{\rm{th}}$ in
this interval becomes smaller than the cold inflation value (dotted
curve).   However, for $Q_{*} \gtrsim 7\times 10^{-5}$, the behavior
is reversed and then  it is noted that the establishment of a FDR is
unfavored in comparison to the cold  inflation case, since higher
values of $\varphi_{\rm{th}}$ demand a higher initial  condition for
inflaton field, $\varphi_0$, for eternal inflation to occur.

When we account for the radiation noise effect (dashed curve), it
becomes relevant only for $Q_{*} \gtrsim 2\times 10^{-4}$ and acts by
increasing even more the value of $\varphi_{\rm{th}}$  in comparison
to the solid curves, thus turning the FDR suppression tendency of warm
inflation stronger.  A simple reasoning about this behavior can be
obtained analyzing Eqs.~(\ref{newP})  and (\ref{newP_RADNOISE}). Since
the radiation noise contributes with a multiplicative  factor
$\mathcal{O}(40)$ to the dissipative power spectrum, the effects
inherent to  warm inflation (FDR suppression which manifests at larger
values for the dissipation and thus, damping effects are stronger)
are expected to be enhanced.  On the other hand, in the case where
$n_k$ is negligible (dash-dotted curve),  for $Q_{*} \gtrsim 10^{-12}$
a FDR is more favored than in cold inflation,  being more prominent at
$Q_{*}\approx10^{-2}$.  Therefore, comparing the results shown by the
dash-dotted and solid curves, one notices the deleterious  role of the
inflaton thermalization to the establishment of a FDR.

{}For the quartic potential ($n=2$), shown in panel (b) of
{}Fig.~\ref{phith_Qstar}, one observes the same qualitative behavior
of the quadratic case. {}For $Q_{*}$ approximately between $10^{-10}$
and $1\times10^{-6}$ (inset plot), a FDR is favored in comparison to
cold inflation, whereas for higher values of $Q_*$ the presence  of a
FDR is harder to be achieved.  
Like in the quadratic case, the effect of the radiation noise on the
power spectrum makes a FDR  even harder to be achieved when $Q_{*}
\gtrsim 10^{-4}$ and the effect of a negligible $n_k$  has the same
FDR favoring behavior.

{}The sextic potential ($n=3$) case, shown in panel (c) of
{}Fig.~\ref{phith_Qstar}, does not favor a FDR for very small $Q_{*}$
like we have seen for the quadratic and quartic cases. The occurrence
of a FDR is always disfavored for $Q_{*}\gtrsim 10^{-8}$, whereas for
lower $Q_*$ the values of $\varphi_{\rm{th}}$ does not fall bellow the
cold inflation one.  However, the qualitative behavior of
$\varphi_{\rm{th}}$ due to the effects of radiation noise and
negligible  $n_k$ are exactly the same of the aforementioned
potentials.  

Due to the qualitative similarity of the dependencies of
$\varphi_{\rm{th}}$ and $N_{\rm{th}}$ on $Q_*$,  we choose to present
plots only for the former and obtain a semianalytic approximation for
the functional  dependence of $N_{\rm{th}}$ on $\varphi_{\rm{th}}$,
which can be found to be well approximated by the expression  

\begin{align} \label{Nth_CH_analytic}
& N_{\rm{th}} =
  \frac{1}{4n}\left(\frac{\varphi_{\rm{th}}}{M_p}\right)^2 \;.
\end{align} 
This solution was obtained by integrating Eqs.~(\ref{dphidN_chaotic})
and (\ref{dQdN_chaotic}) analytically and  inspecting the dominant
terms. This result shows that $N_{\rm{th}}$ possesses the same
qualitative behavior of  $\varphi_{\rm{th}}$ with respect to making a
FDR easier or harder to be achieved due to the combined effects of
dissipation and thermal radiation.  This means that the higher the
value of $\varphi_{\rm{th}}$ we need for eternal inflation to occur,
the larger is  the number of {\it e}-folds of inflation required to
accomplish it and vice versa.  Although Eq.~(\ref{Nth_CH_analytic})
does not contain any explicit dissipative or thermal variable, the
calculation  of the values of $\varphi_{\rm{th}}$ already incorporate
these effects.

%%%%%%%%%%%%%%%%%%%%%%%%%%%%%     FIG   %%%%%%%%%%%%%%%%%%%%%%%%%%%%%%%

\begin{figure}[htb!]
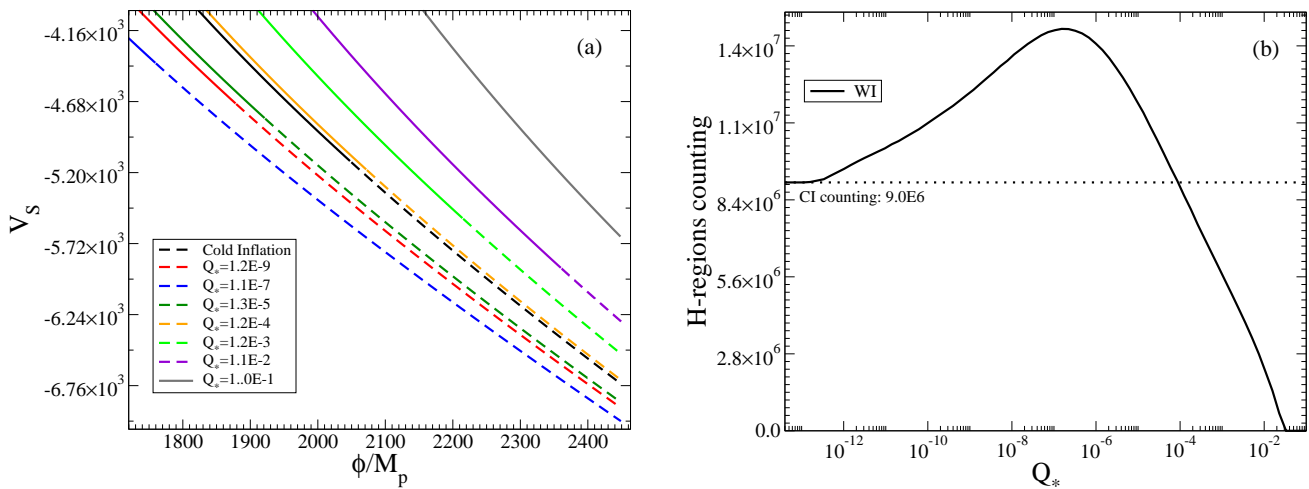

\vspace{0.75cm} \centerline{
  \psfig{file=fig6a.eps,scale=0.3,angle=0}
\hspace{0.5cm} \psfig{file=fig6b.eps,scale=0.3,angle=0}
}   
\caption{The effective potential $V_S$ as a function of
  $x=\phi/M_p$, panel (a), for some representative values of $Q_{*}$
  and the counting of H regions versus $Q_{*}$, panel (b), for the
  quadratic chaotic inflation potential.  It was taken $N_e=1.5\times{10}^6$
  for the cold inflation case ($Q=0$).}
\label{VS_n1}
\end{figure}

\begin{figure}[htb!]
\vspace{0.75cm} \centerline{
  \psfig{file=fig7a.eps,scale=0.3,angle=0}
\hspace{0.5cm} \psfig{file=fig7b.eps,scale=0.3,angle=0}
}
\caption{The effective potential $V_S$ as a function of
$x=\phi/M_p$, panel (a), for some representative values of $Q_{*}$
and the counting of H regions versus $Q_{*}$, panel (b), for the
quartic chaotic inflation potential. It was taken $N_e=4\times{10}^4$ for
the cold inflation case ($Q=0$).}
\label{VS_n2}
\end{figure}

\begin{figure}[htb!]
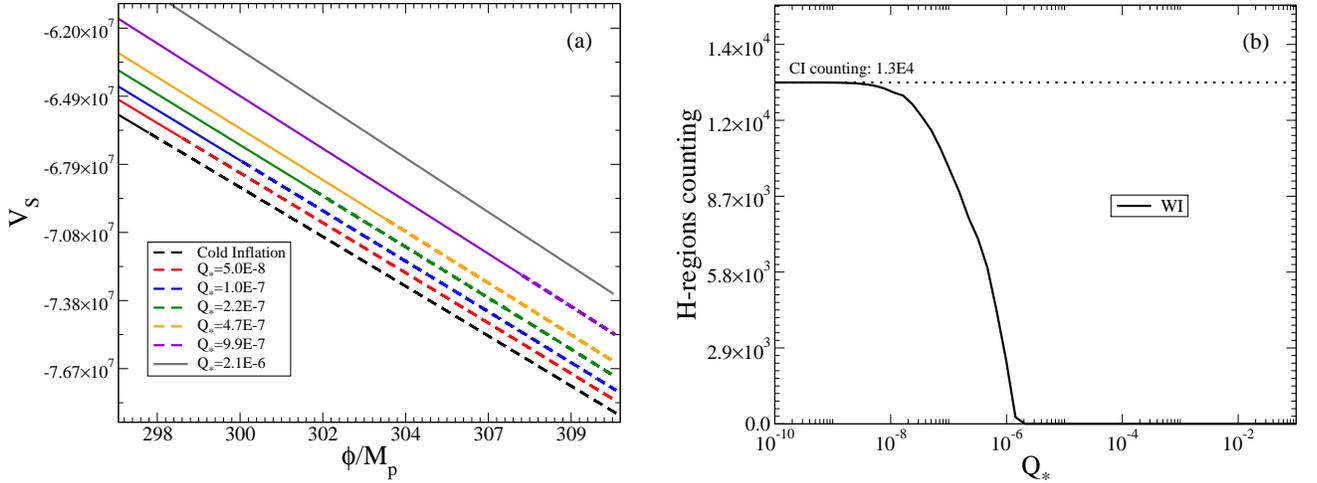

\vspace{0.75cm} \centerline{
  \psfig{file=fig8a.eps,scale=0.3,angle=0}
\hspace{0.5cm} \psfig{file=fig8b.eps,scale=0.3,angle=0}
}
\caption{The effective potential $V_S$ as a function of
$x=\phi/M_p$, panel (a), for some representative values of $Q_{*}$
and the counting of H regions versus $Q_{*}$, panel (b), for the
sextic chaotic inflation potential. It was taken $N_e=8\times{10}^3$ for
the cold inflation case ($Q=0$).}
\label{VS_n3}
\end{figure}

Next, we present in {}Figs.~\ref{VS_n1}, \ref{VS_n2}, and \ref{VS_n3},
the results for the effective potential  $V_S$ as a function of the
dimensionless inflaton field [panel (a) in each of the figures] for
some representative values of $Q_*$. We also show in parallel [panel
(b) in each of the figures], the corresponding counting of H regions
as a function of $Q_*$,  for each of the chaotic inflation  potential
models considered. 
{}For each pair of plots for $V_S$ and H regions counting, we have set
up an adequate initial condition $\varphi_0$  for the cold inflation
case such that a FDR is present.  This same initial condition
$\varphi_0$ was then used to obtain all warm inflation curves of $V_S$
and for  each point of the plots of counting of H regions.  {}From
this perspective, of same value for $\varphi_0$ for both cold and warm
inflation cases, one can inspect whether  the FDR generated in the
cold inflation case is still sustained or becomes suppressed when
dissipative effects are present.  In the plots of $V_S$, the curves
are separated in FDR and deterministic parts as described in the
previous  subsection for the cold inflation situation. One notices
that the lengths of the parts corresponding to FDR increase or
decrease due to the dependence of $\varphi_{\rm{th}}$ on dissipation
and temperature  [see {}Figs.~(\ref{vstar_Qstar}] and
(\ref{phith_Qstar})), thus revealing the enhancement or suppression
of the FDR for each representative value of $Q_*$.  In turn, the plots
of H regions counting exhibit the number of H regions produced in the
FDR parts shown  in the plots for $V_S$.

{}For the quadratic potential case, shown in {}Fig.~(\ref{VS_n1}), we 
observe in panel (a) that the FDR dash-dotted lines increase until 
$Q_{*}\approx 1.1\times 10^{-7}$, which reveals a favoring tendency to eternal 
inflation in comparison  to cold inflation.
Increasing $Q_{*}$, this behavior is reversed and for $Q_{*}\gtrsim 1\times10^{-4}$, 
eternal inflation is disfavored. 
Panel (b) corroborates this behavior in terms
of the increase and subsequent decrease  of the production of
H regions. 
The corresponding value of temperature (at that particular time) for which 
the production of H regions decreases, i.e., eternal inflation gets disfavored,
is $T_{\rm{th}}\gtrsim 1.6\times 10^7$ GeV.
One also notes that the counting of H regions falls to zero
for $Q_*\gtrsim 3\times 10^{-2}$. The solid grey curve in panel (a) shows an example
where no FDR is present.
This fall means that the chosen $\varphi_0$ of cold
inflation is not above the threshold value to produce a FDR in warm
inflation with such values of $Q_*$.  This result is in complete
consistency with the ones shown in {}Fig.~\ref{phith_Qstar}(a).

The quartic potential case, shown in {}Fig.~(\ref{VS_n2}), is
qualitatively similar to the quadratic case. 
Panel (a)  shows that a FDR is more favored than in cold inflation for 
values $Q_*$ between $1.8\times 10^{-9}$ and  $6.8\times 10^{-7}$, but 
for higher values of $Q_*$ the tendency is the suppression of the FDR.
In panel (b), we ignore the mentioned negligible FDR favoring for 
very low $Q_*$ (no inset plot) and reiterate that for $Q_*\gtrsim 1\times 10^{-6}$
eternal inflation is disfavored in comparison to cold inflation as the
solid curve drops below cold inflation dotted one. The corresponding value of 
temperature in this case is $T_{\rm{th}}\gtrsim 9.1\times 10^{10}$ GeV.
{}For the chosen value of $\varphi_0$, for $Q_*\gtrsim 4\times 10^{-6}$,  eternal
inflation is completely suppressed.  This result also corroborates the
one shown in {}Fig.~\ref{phith_Qstar}(b).

The case of a sextic potential, shown in {}Fig.~(\ref{VS_n3}), 
panel (a) shows that 
the FDR is always disfavored as we increase $Q_*$.  {}For $Q_*\gtrsim
5\times 10^{-8}$, the lengths of the FDR curves becomes smaller in comparison
to the cold inflation case until  it disappears for $Q_*\approx
2\times 10^{-6}$. The exactly same behavior is shown in panel (b),
where the production of H regions falls below the cold inflation value
for corresponding values of temperature of $T_{\rm{th}}\gtrsim 3.9\times 10^{12}$ GeV,
and eventually is totally suppressed.   These results are again
consistent to those shown in {}Fig.~\ref{phith_Qstar}(c).

With the assistance of {}Fig.~(\ref{vstar_Qstar}), we notice that the
FDR favoring intervals of $Q_*$ obtained in the quadratic and  quartic
cases occur for $T/H\lesssim 1$, which is a regime between cold and
warm inflation regimes, which we called WWI.  {}For the typical warm
inflation picture (where $T/H\gtrsim 1$), one observes that
dissipation has the tendency to suppress the establishment of a SRR
for the case of the chaotic potential models in warm inflation, in
comparison to the  cold inflation case.

%%%%%%%%%%%%%%%%%%%%%%%%%%%%%%%%%%%%%%%%%%%%%%%%%%%%%%%%%%%%%%%%%%%%%%
\subsection{Hilltop warm inflation case}
\label{hilltop}

We now discuss and present our results for the hilltop potential case,
given by Eq.~(\ref{potential_hybrid_nneq0}). Equation~(\ref{EoM}) can
be specialized in order to describe the inflaton dynamics under this
potential

\begin{eqnarray}  \label{cold_EoM_hilltop}
\frac{\partial x}{\partial t'}  &=&
\frac{|\gamma|}{2n}\frac{x^{2n-1}}{\sqrt{1-\left(|\gamma|/2n\right)
    x^{2n}}(1+Q)} \nonumber\\ &+& \frac{\sqrt{3\lambda}}{6n}
\frac{\left[1-\left(|\gamma|/2n\right)x^{2n}\right]^{3/4}}{2\pi}
\left(1+2n_{\tilde k}\right) \zeta_q'  \nonumber\\ &+&
\frac{\sqrt{3\lambda}}{6n}
\frac{\left[1-\left(|\gamma|/2n\right)x^{2n}\right]^{3/4}}{2\pi}
\left\{\frac{T'}{\left[1-\left(|\gamma|/2n\right)x^{2n}\right]^{1/4}}
\right\}\frac{\pi Q}{10(1+Q)^2} \zeta_T'\;.
\end{eqnarray}
The modified Fokker-Planck equation for the volume distribution $P_V$,
Eq.~(\ref{FPdim}), can be written as 

\begin{eqnarray} \label{warmFP_small_field}
&& \frac{\partial }{\partial t'} P_V(x ,t') = -\frac{\partial
  }{\partial x}
  \left[\frac{|\gamma|}{2n}\frac{x^{2n-1}}{\sqrt{1-\left(|\gamma|/2n\right)
        x^{2n}}(1+Q)} P_V(x,t') \right] \nonumber\\ &+&
  \frac{\partial^2 }{\partial x^2} \left\{ \frac{\lambda}{12n^2}
  \frac{\left[1-\left(|\gamma|/2n\right)x^{2n}\right]^{3/2}}{8\pi^2}
  \left[1+2n_{\tilde k} +
    \left(\frac{T'}{\sqrt{1-\left(|\gamma|/2n\right)x^{2n}}}\right)\frac{\pi
      Q}{10(1+Q)^2}\right]  P_V(x,t') \right\}
  \nonumber\\ &+&\frac{3}{2n}\sqrt{1-\left(|\gamma|/2n\right)x^{2n}}
  P_V(x,t') \;. 
\end{eqnarray}

{}For the analysis of the hilltop case, we set two values of $\gamma$
for each fixed $n$. Namely, we set $\gamma=10^{-3}$ and
$\gamma=10^{-2}$ for $n=1$, and $\gamma=10^{-5}$ and $\gamma=10^{-4}$
for $n=2$. These values of $\gamma$ are motivated by those values
considered in the recent Planck's observational constraints on
inflation based on the hilltop  potential~\cite{Planck2015} [note also
that in~\cite{Planck2015}, these values were given in terms of
$\log{(\mu/M_p)}$ instead].

%%%%%%%%%%%%%%%%%%%%%%%%%%%% FIG %%%%%%%%%%%%%%%%%%%%%%%%%%%%%%%%%%%%%%%%

\begin{figure}[htb!]
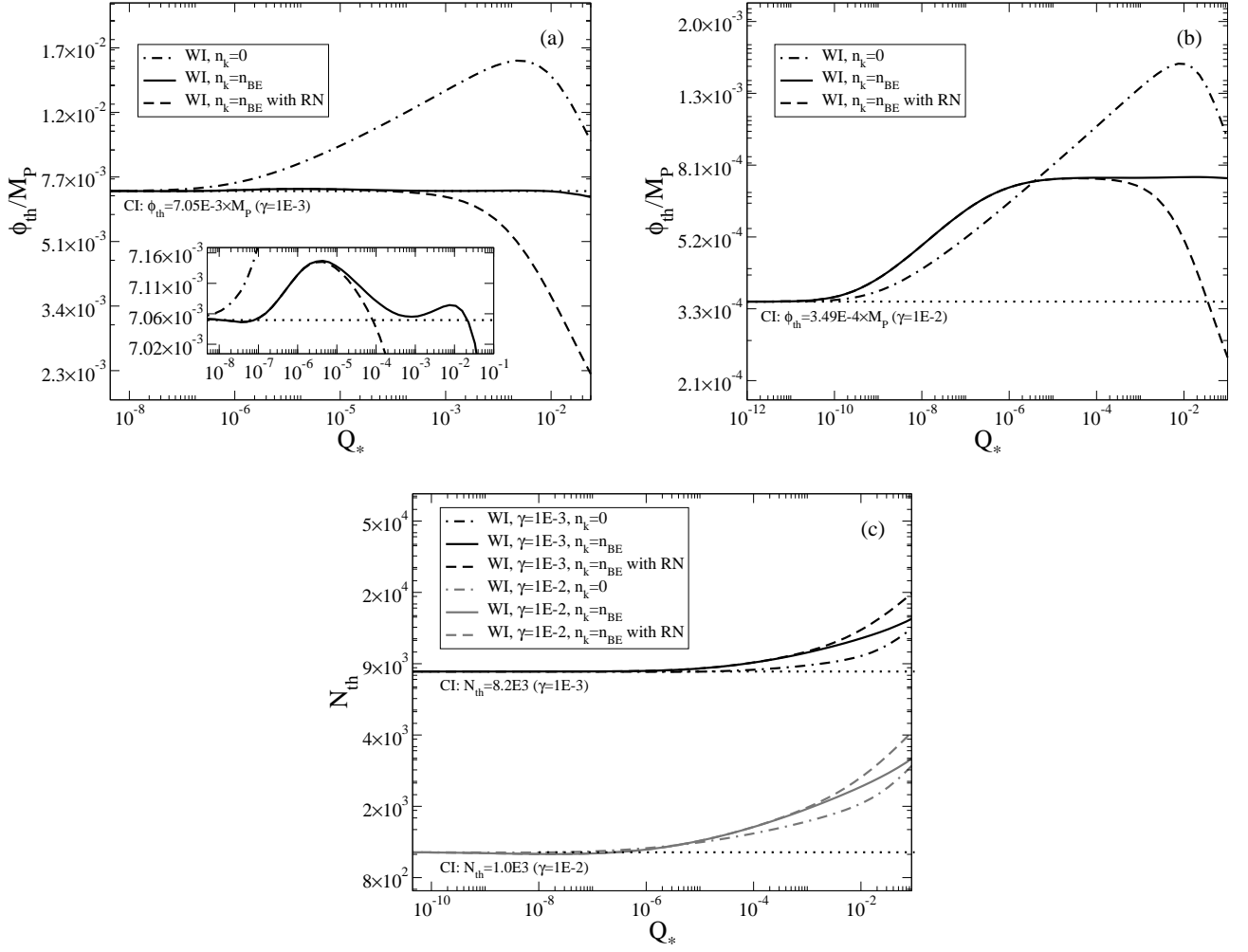

\vspace{0.75cm} \centerline{
  \psfig{file=fig9a.eps,scale=0.30,angle=0}
\hspace{0.5cm}
\psfig{file=fig9b.eps,scale=0.30,angle=0} }
\vspace{0.5cm}  \centerline{
  \psfig{file=fig9c.eps,scale=0.30,angle=0}}
\caption{ The threshold values of $\phi_{\rm{th}}$ [panels (a) and
  (b)] and $N_{\rm{th}}$ [panel (c)] versus $Q_{*}$ for the quadratic
  hilltop inflation potential case. Panels (a) and (b) correspond to
  the representative choices $\gamma=10^{-3}$ and $\gamma=10^{-2}$,
  respectively, whereas panel (c)  covers both $\gamma=10^{-3}$ (black
  dashed curves) and $\gamma=10^{-2}$ (gray dashed curves) choices.
  The solid (dash-dot) curves correspond to thermal (negligible)
  inflaton distribution $n_k$, while the dashed curve corresponds to
  thermal inflaton distribution accounting for radiation noise
  contribution.}
\label{Nth_Qstar_HT_n1}
\end{figure}

\begin{figure}[htb!]
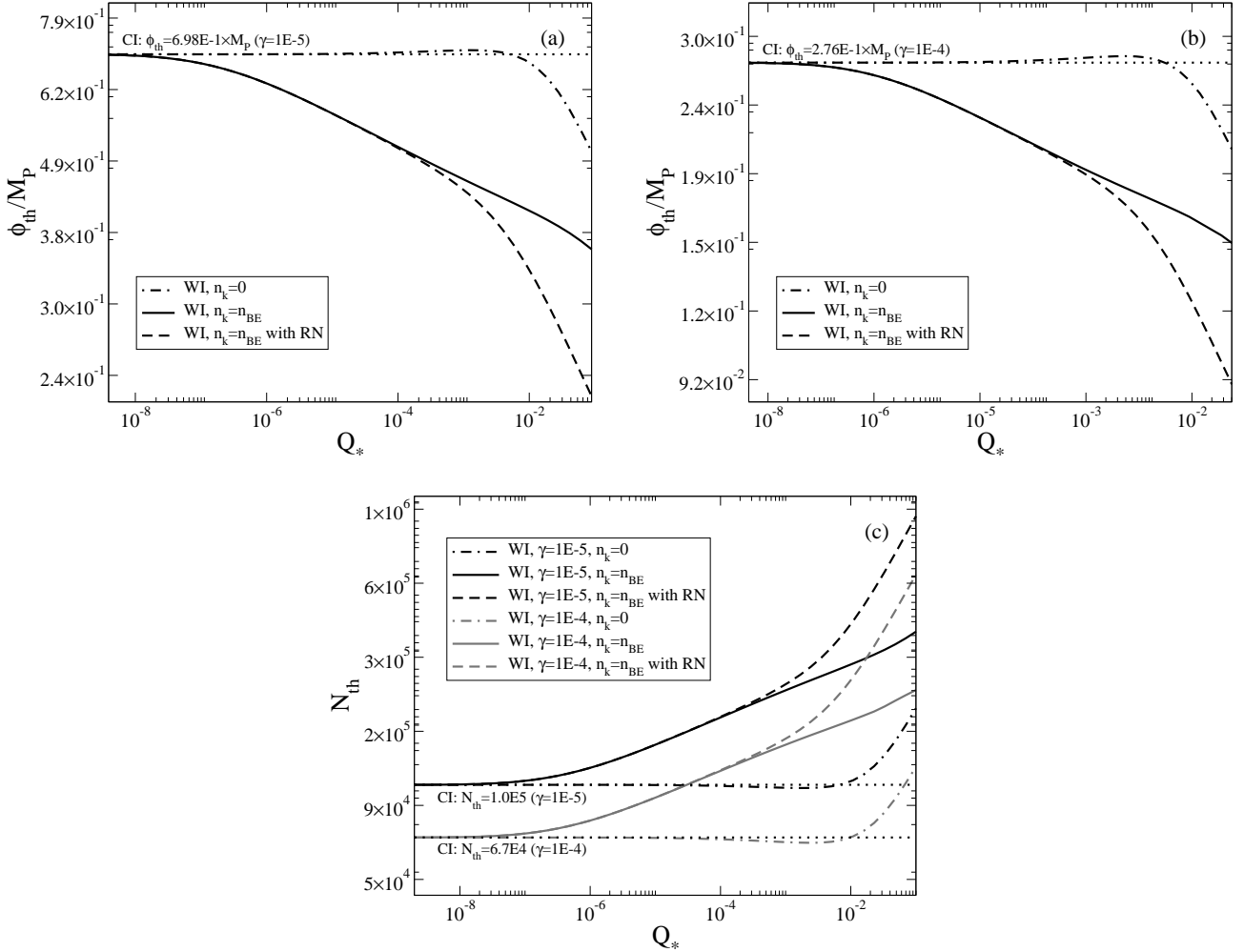

\vspace{0.75cm} \centerline{
  \psfig{file=fig10a.eps,scale=0.30,angle=0}
\hspace{0.5cm}
\psfig{file=fig10b.eps,scale=0.30,angle=0} }
\vspace{0.5cm}  \centerline{
  \psfig{file=fig10c.eps,scale=0.30,angle=0}}
\caption{ The same as in {}Fig.~\ref{Nth_Qstar_HT_n1}, but for the
  quartic hilltop inflation potential case.  Panels (a) and (b)
  correspond to the representative choices $\gamma=10^{-5}$ and
  $\gamma=10^{-4}$, respectively, whereas panel (c)  covers both
  $\gamma=10^{-5}$ (black curves) and $\gamma=10^{-4}$ (gray curves)
  choices.  The solid (dash-dot) curves correspond to thermal
  (negligible) inflaton distribution $n_k$, while the dashed curve
  corresponds to thermal inflaton distribution accounting for
  radiation noise contribution.}
\label{Nth_Qstar_HT_n2}
\end{figure}

In {}Figs.~\ref{Nth_Qstar_HT_n1} and \ref{Nth_Qstar_HT_n2}, we present
the functional dependence  of $\varphi_{\rm{th}}$ and $N_{\rm{th}}$ on
the dissipation ratio $Q_*$ for the hilltop potential model cases.
Differently to what we have seen in the chaotic potential case, for
hilltop inflation the relation between $\varphi_{\rm{th}}$  and
$N_{\rm{th}}$ is much more involved. Thus, we show the numerical
results for both in this case.  The numerical results for
$\varphi_{\rm{th}}$ and $N_{\rm{th}}$ as a function of $Q_*$ are shown
in  {}Figs.~\ref{Nth_Qstar_HT_n1} and \ref{Nth_Qstar_HT_n2} for the
quadratic and for the quartic hilltop inflation potential cases,
respectively.  Panels (a) and (b) of each figure show the functional
dependence of $\varphi_{\rm{th}}$ on $Q_*$ for each aforementioned
choice of $\gamma$, whereas panel (c) shows the functional  dependence
of $N_{\rm{th}}$ on $Q_*$.  Note that for sufficiently small values
for $Q_*$, the cold inflation limit is recovered in all panels, as
expected.

The quadratic inflation case for $\gamma=10^{-3}$ is shown in panel
(a) of Fig.~\ref{Nth_Qstar_HT_n1}.  One observes that as we increase
$Q_*$, the value of $\varphi_{\rm{th}}$ is larger than in the cold
inflation  case (which is better seen in the inset plot). {}For
hilltop inflation potential this means that the SRR is favored.  
In other words, since for given values of $Q_*$ the amplitude of the
inflaton increases, i.e., moves away from the top of the potential,
the region of field values between cold and warm inflation values of
$\varphi_{\rm{th}}$ become now available for the SRR. 
Consequently, we see that a larger region of field values in
warm inflation becomes suitable for leading to eternal inflation than
in the cold inflation case.    This favoring occurs for $Q_*\gtrsim
10^{-7}$ and is more pronounced at $Q_*\approx 4 \times 10^{-6}$ and
$Q_*\approx10^{-2}$.  However, for $Q_*\gtrsim 10^{-2}$, the behavior
is reversed and the FDR tends to be unfavored for 
$Q_*\gtrsim 2 \times 10^{-2}$.
This same FDR friendly behavior happens for the
quadratic case with $\gamma=10^{-2}$, shown in panel (b), which occurs
for $Q_*\gtrsim 4 \times 10^{-11}$
and stabilizes for $Q_*\gtrsim 5 \times 10^{-5}$.
In contrast, panel (c) reveals that the threshold number of {\it e}-folds
$N_{\rm{th}}$ increases with dissipation for both  choices of
$\gamma$, which indicates that the establishment of a SRR is harder to
be achieved for higher values of $Q_*$.  These results involving
$\varphi_{\rm{th}}$ and $N_{\rm{th}}$ seem contradictory to the ones
seen for the chaotic  inflation potential cases, where we would expect
growing $\varphi_{\rm{th}}$ for growing $N_{\rm{th}}$ and vice versa.
However, this apparent contradiction can be dissolved when we realize
that at the same time that dissipative effects become sufficiently
significant at the threshold instant to increase the values of
$\varphi_{\rm{th}}$,  the inflaton field value at the end of
inflation, $\varphi_{\rm{f}}$, becomes smaller due to dissipation,
thus increasing $N_{\rm{th}}$.

The quartic inflation cases for $\gamma=10^{-5}$ and $\gamma=10^{-4}$
are shown in panel (a) and (b) of  {}Fig.~\ref{Nth_Qstar_HT_n2},
respectively. {}For both cases, eternal inflation is continuously
suppressed as we increase $Q_*$ from approximately $10^{-8}$ to
greater values.  These behaviors are in agreement with the respective
results of $N_{\rm{th}}$ given in panel (c), since we now expect that
in the hilltop inflation potential lower values of $\varphi_{\rm{th}}$
will disfavor a FDR, which corresponds to greater  values of
$N_{\rm{th}}$. 

In both {}Figs.~\ref{Nth_Qstar_HT_n1} and \ref{Nth_Qstar_HT_n2}, we
also analyze the effect of a negligible  inflaton particle
distribution $n_k \approx 0$. In {}Figs.~\ref{Nth_Qstar_HT_n1} and
\ref{Nth_Qstar_HT_n2}, this case is represented by dash-dotted lines
in all panels. The behavior is similar to the chaotic inflation case;
the curves ascend in comparison to the $n_k=n_{\rm{BE}}$ cases (solid
curves), which again reinforces the importance of the thermalized
inflation particles in the suppression of eternal inflation. In the
quadratic inflation case, the establishment of a SRR is always
favored, whereas in the quartic case eternal inflation is  negligibly
favored for very low values of $Q_*$ and becomes significantly
suppressed for $Q_*\gtrsim 10^{-2}$.
In both figures, we also see the
effect of radiation noise contribution to the power spectrum, given
according to Eq.~(\ref{newP_RADNOISE}). Its effect is opposite to that
of negligible $n_k$; in the quadratic case with $\gamma = 10^{-3}$,
the curves descend from $n_k=n_{\rm{BE}}$ case for  $Q_*\gtrsim
5 \times 10^{-5}$, whereas for the quadratic case with $\gamma= 10^{-2}$ and
for both quartic values of $\gamma$ in  the quartic potential the
descent happens for $Q_*\gtrsim 10^{-4}$. In the quadratic case with
$\gamma =10^{-3}$  ($\gamma= 10^{-2}$), eternal inflation is favored
for $Q_*\gtrsim 10^{-4}$ ($Q_*\gtrsim 3\times10^{-2}$), whereas in
the quartic case eternal inflation is always favored for $Q_*\gtrsim
10^{-4}$. This effect of the radiation noise is consistent to what we
expected before in the chaotic inflation potential cases. The effect
of the radiation noise is more pronounced at larger dissipation. 
These larger values of dissipation imply in a larger 
damping of fluctuations that might otherwise lead to a SRR.

%%%%%%%%%%%%%%%%%%%%%%%%%%%%%%%%%%%%%%%    FIG  %%%%%%%%%%%%%%%%%%%%%
\begin{figure}[htb!]
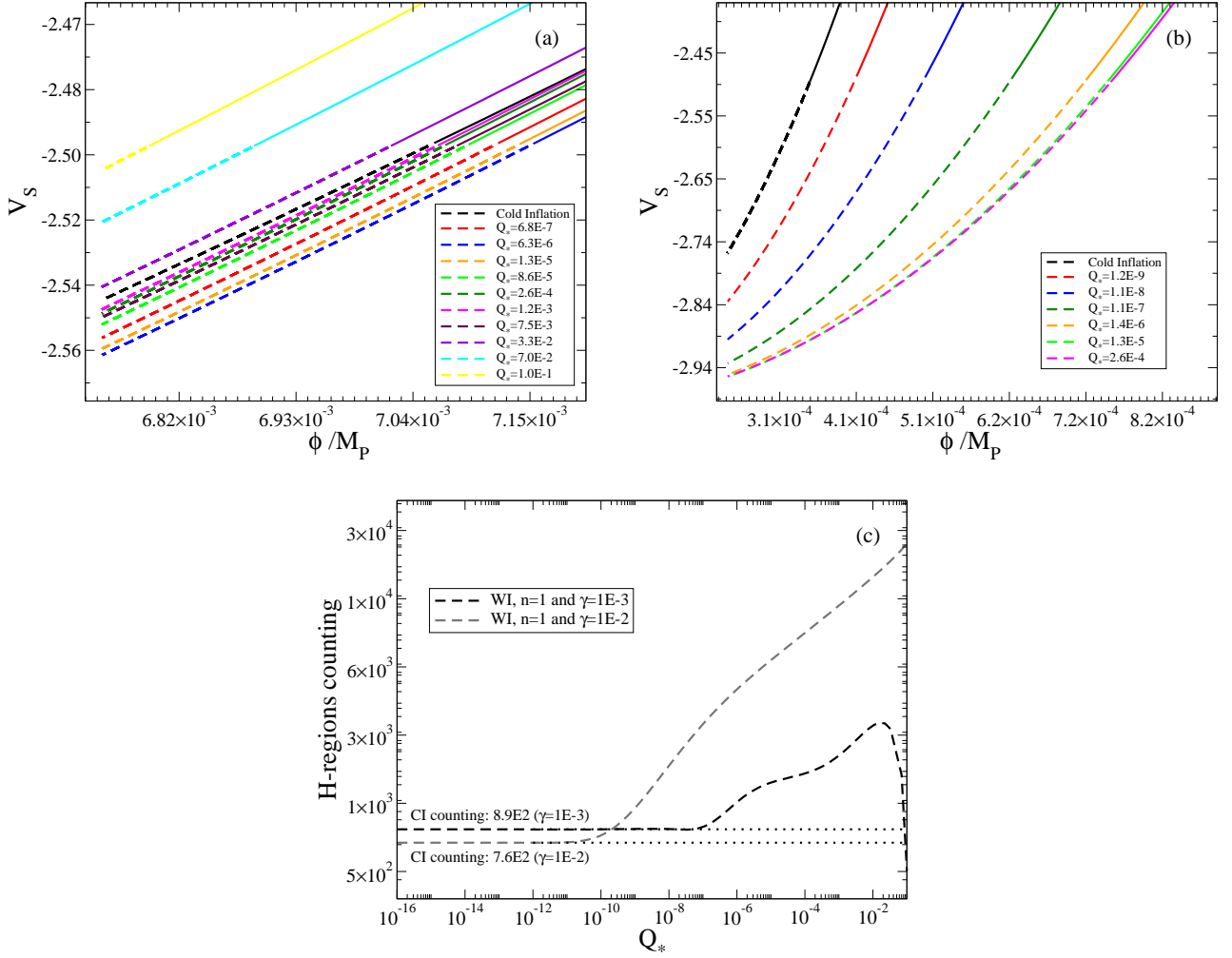

\vspace{0.75cm} \centerline{
  \psfig{file=fig11a.eps,scale=0.3,angle=0}
\hspace{0.5cm}
\psfig{file=fig11b.eps,scale=0.3,angle=0} }
\vspace{0.5cm}  \centerline{
  \psfig{file=fig11c.eps,scale=0.3,angle=0} }
\caption{The effective potential $V_S$ as a function
  of $\phi$ for some representative values of $Q_{*}$  [panels (a) and
  (b)] and the counting of H regions versus $Q_{*}$ [panel (c)], for
  the quadratic hilltop inflation potential case.  Panels (a) and (b)
  correspond to the representative choices of $\gamma=10^{-3}$ and
  $\gamma=10^{-2}$, respectively,  whereas panel (c) covers both
  $\gamma=10^{-3}$ (black curve) and $\gamma=10^{-2}$ (gray curve)
  choices.  We have chosen $N_e=8300$ ($\gamma=10^{-3}$) and
  $N_e=1050$ ($\gamma=10^{-2}$) for cold inflation cases.}  
\label{VS_HT_n1}
\end{figure}

\begin{figure}[htb!]
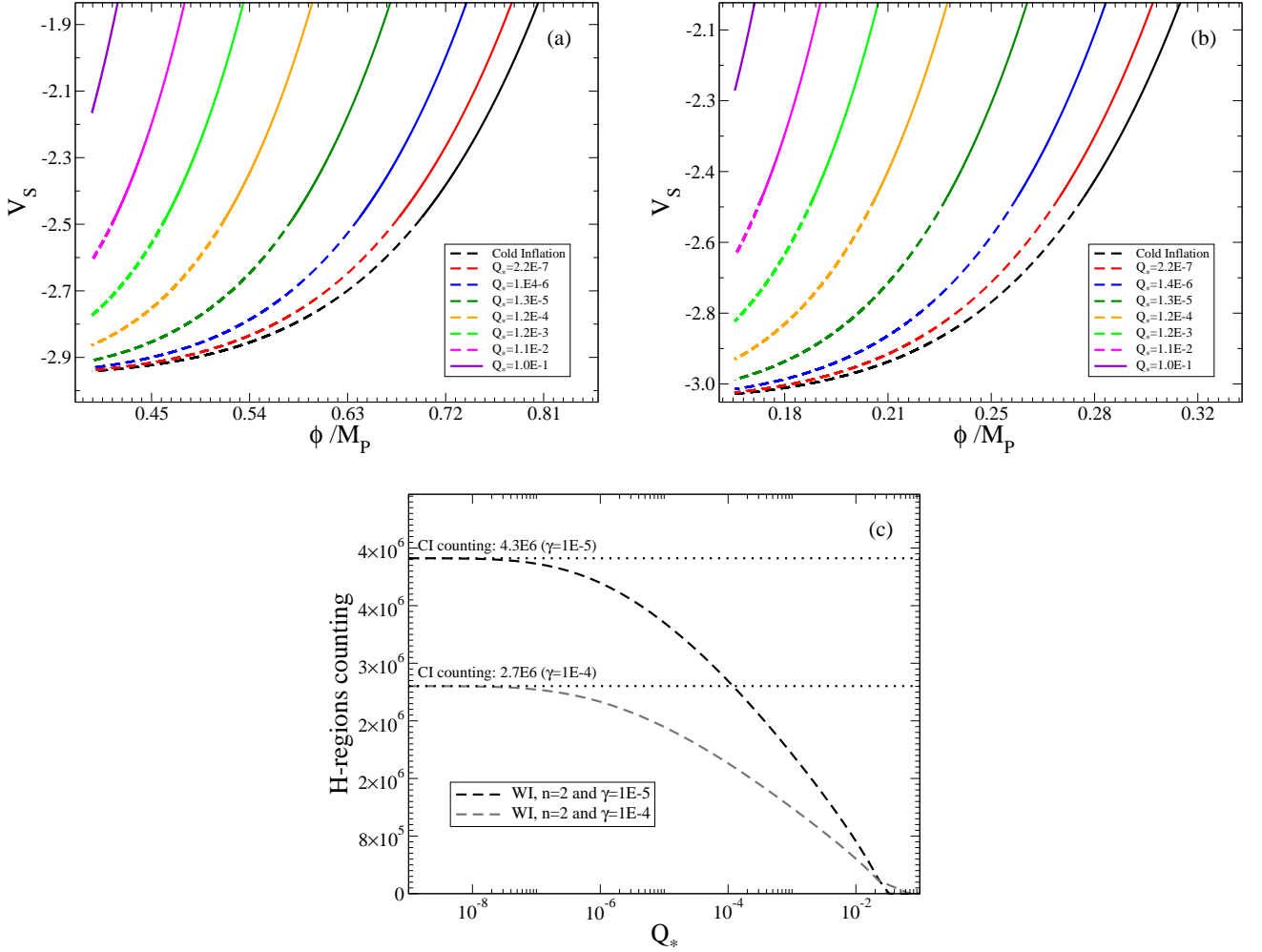

\vspace{0.75cm} \centerline{
  \psfig{file=fig12a.eps,scale=0.3,angle=0}
\hspace{0.5cm}
\psfig{file=fig12b.eps,scale=0.3,angle=0} }
\vspace{0.5cm}  \centerline{
  \psfig{file=fig12c.eps,scale=0.3,angle=0} }
\caption{The same as in {}Fig.~\ref{VS_HT_n1}, but for
  the quartic hilltop inflation potential case.  Panels (a) and (b)
  correspond to the representative choices $\gamma=10^{-5}$ and
  $\gamma=10^{-4}$, respectively,  whereas panel (c) covers both
  $\gamma=10^{-5}$ (black curve) and $\gamma=10^{-4}$ (gray curve)
  choices.  We have chosen $N_e=3.2\times10^5$ ($\gamma=10^{-5}$) and
  $N_e=2.0\times10^5$ ($\gamma=10^{-4}$) for cold inflation cases.}    
\label{VS_HT_n2}
\end{figure}

{}Finally, in {}Figs.~\ref{VS_HT_n1} and \ref{VS_HT_n2}, we present the
effective potential $V_S$ as a function of the (dimensionless)
inflaton field for some representative values of $Q_*$ [panels (a) and
(b)] and the functional dependence of the counting of H regions on
$Q_*$ [panel (c)], for the quadratic and quartic hilltop potentials,
respectively.  Panels (a) and (b) of each figure show the plots of
$V_S$ for each choice of $\gamma$, whereas panels (c)  exhibit the
counting of H regions for both choices of $\gamma$.  As in the chaotic
inflation case, we choose suitable values of $\varphi_0$ for the cold
inflation cases such that a FDR is present and use it to obtain the
warm inflation results.  We can now contrast the panel (a)  of
{}Fig.~\ref{VS_HT_n1} with the corresponding panel (a) of
{}Fig.~\ref{Nth_Qstar_HT_n1}.
Observing both
panels, one notes that the FDR is favored in the range $10^{-7}
\lesssim Q_*\lesssim 2 \times 10^{-2}$ in comparison to the cold 
inflation case and unfavored outside this range. Quantitatively, the 
region of the potential that corresponds to the FDR (i) increases for 
$10^{-7}\lesssim Q_*\lesssim 4 \times 10^{-6}$, (ii) decreases in the range 
$ 4 \times 10^{-6}\lesssim Q_*\lesssim 8 \times 10^{-4}$, (iii) increases for 
$ 8 \times 10^{-4} \lesssim Q_*\lesssim 8 \times 10^{-3}$, and (iv) decreases 
for $Q_* \gtrsim 8 \times 10^{-3} $, getting shorter than the cold inflation 
case for $Q_* \gtrsim 2 \times 10^{-2}$.
Analogously, we perform a joint analysis of panels (b) of
{}Figs.~\ref{Nth_Qstar_HT_n1} and \ref{VS_HT_n1}.  {}For $Q_*\gtrsim
4 \times 10^{-11}$ in both panels (b), one observes that the FDR abruptly
increases for increasing $Q_*$ until $Q_*\approx 10^{-6}$, where FDR
continuous to increase but in a small rate.  These minor details
involving representative values of $Q_*$ are important only to
contrast the corresponding panels of  {}Figs.~\ref{Nth_Qstar_HT_n1}
and \ref{VS_HT_n1}, but the main interest is in the FDR-favoring
behavior that we observe for the quadratic  hilltop potential case.
Analogously, we contrast panels (a) and (b) of {}Fig.~\ref{VS_HT_n2}
to the respective panels (a) and (b) from
{}Fig.~\ref{Nth_Qstar_HT_n2}.  One notices that for both cases the
lengths of the FDR curves decrease until it disappear for sufficiently
high value of $Q_*$, which corresponds to the decrease of the values
of $\varphi_{\rm{th}}$.  The plots of the counting of H regions, given
in the panels (c) of {}Figs.~\ref{VS_HT_n1} and ~\ref{VS_HT_n2}, mimic
the  results of panels (a) and (b) like in the chaotic inflation case;
the amount of H regions increases when FDR is favored, decreases when
it is unfavored, and falls to zero when the chosen $\varphi_0$ is
smaller than $\varphi_{\rm{th}}$ for the  specific value of
$Q_*$. Like in the cases of the monomial chaotic potential, the 
corresponding values of temperature for which 
eternal inflation gets disfavored in comparison to cold inflation are
$T_{\rm{th}}\gtrsim 4.0\times 10^{12}$ GeV for the quadratic case with $\gamma=10^{-3}$
and $T_{\rm{th}}\gtrsim 2.6\times 10^{7}$ GeV ($T_{\rm{th}}\gtrsim 1.6\times 10^{7}$ GeV) 
for the quartic case with $\gamma=10^{-5}$ ($\gamma=10^{-4}$),
whereas for the quadratic hilltop case with $\gamma=10^{-2}$ eternal inflation
is favored for the entire range of $Q_*$ analyzed.
Once again, these plots reveal the deleterious behavior of warm
inflation to the establishment of the SRR in the quartic hilltop
potential case. In the quadratic hilltop potential case, however, warm
inflation enhances the eternal  inflation mechanism from the point of
view of a fixed $\varphi_0$.

%%%%%%%%%%%%%%%%%%%%%%%%%%%%%%%%%%%%%%%%%%%%%%%%%%%%%%%%%%%%%%%%%%%%%%
\section{Conclusions and final remarks}
\label{sec6}

In this work, we have developed a generalized approach to eternal
inflation of the random walk type under the framework of  warm
inflation.  Thus, the combined effects of dissipation, the
corresponding stochastic term, and the presence of a thermal radiation
bath are accounted for.  To understand the influence of these effects
on the self-reproduction regime of the inflationary universe, we have
performed a comprehensive numerical analysis of how relevant
quantities that characterizes eternal inflation in the cold inflation
case are modified due to the presence of dissipation and a thermal
radiation bath. Since eternal inflation is mainly characterized by the
presence of a FDR, the main tool we have used was a generalized
condition for the occurrence of a FDR, which is used for obtaining and
interpreting the results.

Taking cold inflation as a reference, within the context of the warm
inflation picture we have obtained information about the functional
relation  between the threshold inflaton field $\varphi_{\rm{th}}$ and
also for the threshold number of {\it e}-folds $N_{\rm{th}}$, in terms of
the dissipation ratio $Q$ (where we used its value at the moment of
horizon crossing as a reference).  In addition to the usual case where
the statistical occupation number is given  by the Bose-Einstein
distribution, i.e., assuming a thermal equilibrium distribution for
the inflaton, we have also presented the cases where its particle
distribution is negligible and where the dissipative power spectrum
might get additional contributions due to radiation noise effects, as
recently studied in Ref.~\cite{bastero2014cosmological}.  Using the
model independent relation between $Q_*$ and $T_*/H_*$, we were able
to focus the analysis only as a function  of $Q_*$, but always
concomitantly keeping track of the influence of the temperature of the
thermal  bath. 

In addition to the analysis of $\varphi_{\rm{th}}$ and $N_{\rm{th}}$,
we have performed a SLA of the corresponding
Fokker-Planck  equation for the probability of having eternal
inflation, translated in terms  of the effective potential $V_S$. In
parallel to the analysis of $V_S$, we have also analyzed the
dependence of the number of H regions produced in the FDR as a
function of $Q_*$.

We have considered as examples of inflation models, the cases of
monomial potentials of the chaotic type (quadratic, quartic,  and
sextic chaotic inflation potentials) and hilltoplike (quadratic and
quartic hilltop  potentials). To study how the typical dynamics
displayed in warm inflation affects eternal inflation, we have
performed our analysis in a range of values for the dissipation ratio
$Q_*$ varying from very low values (reaching the cold inflation limit)
up to $Q_*=0.1$, where the analytical expression for the primordial
spectrum is found to be in good agreement with full numerical
calculations of the perturbations in warm
inflation~\cite{bastero2014cosmological}.

{}For the chaotic potential cases, the dependence of both
$\varphi_{\rm{th}}$  and $N_{\rm{th}}$ on $Q_*$ reveals that in the
typical warm inflation regime dissipation and thermal fluctuations
have the tendency of suppressing eternal  inflation in comparison to
the cold inflation case, whereas in the WWI regime,  eternal inflation
is slightly favored for the quadratic potential case.  When we account
for the radiation noise contribution to the power spectrum, the
suppression tendency becomes even stronger for typical warm inflation
values.  This is expected because, as shown in
Ref.~\cite{bastero2014cosmological}, these  effects become more
relevant for larger values of the dissipation. But this is when
dissipation damps more efficiently the fluctuations that might
otherwise favor  eternal inflation to appear.  However, in the case
where the particle distribution function is negligible,  eternal
inflation effects becomes enhanced for the whole interval of $Q_*$.
This can be traced to the fact that the quantum noise effects have a
larger amplitude, thus favoring the conditions for the emergence of
eternal inflation.

{}For the hilltop potential cases, the dependence of
$\varphi_{\rm{th}}$ on $Q_*$  reveals that eternal inflation is
favored for the quadratic potential and  unfavored for the quartic
potential.  In the case of the quadratic potential, both
$\varphi_{\rm{th}}$ and $N_{\rm{th}}$ grows for increasing $Q_*$,
which means that at the same time dissipation and thermal
fluctuations demand a less restricted value of $\varphi_{\rm{th}}$ for
eternal inflation to happen but, on the other hand, requires a larger
amount of {\it e}-folds for it to take place.  Therefore, depending on the
point of view of fixed $\varphi_0$ or fixed number of {\it e}-folds, eternal
inflation is favored or unfavored, respectively.  In the case of the
quartic potential, $\varphi_{\rm{th}}$ decreases for increasing
$Q_*$, while $N_{\rm{th}}$ increases, which means that for both
point of views described for the quadratic case, both dissipation and
thermal radiation tend to suppress eternal inflation.  
When we account for the radiation noise contribution to the power spectrum, 
the FDR favoring tendency in the quadratic potential is attenuated whereas 
for the quartic potential it turns the suppression tendency stronger for 
sufficiently large values of $Q_*$, which are responsible 
for fluctuation damping. 
When one considers a negligible particle distribution for the inflaton field,
the establishment of a SRR is favored for the whole interval of $Q_*$
in the   quadratic potential case, whereas for the quartic case, FDR
is negligibly  enhanced for very small $Q_*$ and becomes significantly
suppressed for higher $Q_*$.

In summary, our results show that in the chaotic inflation case,
dissipation and thermal fluctuations tend to suppress eternal
inflation in the typical warm inflation dynamics. This
suppression is more pronounced when radiation noise effects on the
power  spectrum are accounted for (which, as already mentioned above,
happens for larger values of dissipation) and eternal inflation
is alleviated when the  statistical distribution of the inflaton is
neglected.  On the other hand, in the hilltop inflation case, for the
quadratic potential, the main tendency is to favor eternal inflation
when we depart from the same $\varphi_0$, but to disfavor it when we
analyze the case where a fixed number of {\it e}-folds is assumed.  In the
quartic case, however, warm inflation effects tend to suppress the SRR
for the whole interval of $Q_*$, which happens for both fixed
$\varphi_0$ or number of {\it e}-folds.  When radiation noise is included,
eternal inflation is even more suppressed for typical warm inflation
values and also suppressed for negligible $n_k$ at sufficiently high
$Q_*$.

Based on the analysis performed, the introduction of warm inflation
effects in the eternal  inflation scenario seems to be deleterious to
the establishment of a self-reproduction regime,  although for some
particular choices of potential and parameters, it is possible to have
exceptions  where eternal inflation is enhanced. This happens
particularly for small values of the dissipation term, in which case,
the fluctuations favoring the presence of a eternal inflation regime
might even be enhanced compared to cold inflation. 
Our results show the nontrivial effects that
dissipation, stochastic noises, and the presence of a thermal
radiation bath, hallmarks of the warm inflation picture, can have in
the global dynamics of inflation and, as studied in this paper, on one
of the most peculiar predictions of the inflationary scenario, eternal
inflation.

%%%%%%%%%%%%%%%%%%%%%%%%%%%%%%%%%%%%%%%%%%%%%%%%%%%%

\appendix

\section{Derivation of $D^{(2)}$} 
\label{D2diss}

The stochastic equation of motion for the inflaton field that involves
both quantum (vacuum) and thermal (dissipative) noises,
Eq.~(\ref{eom}), can be rewritten as 

\begin{equation} \label{eom_appendix}
\dot \varphi = -\frac{V_{,\varphi}}{3H(1+Q)}  +  \eta_q(t) +
\eta_T(t)~,
\end{equation}
where the two-point correlation function for the thermal
noise~\cite{WIreviews1,WIreviews2} is given by

\begin{align}\label{both_correll_appendix}
\langle \eta_T ({\bf x}, t) \eta_T ({\bf x}', t') \rangle = \frac{2  Q
}{3(1+Q)^2}\left(\frac{T}{H}\right) a^{-3} \delta^3({\bf x} - {\bf
  x}') \delta(t-t')~,
\end{align}
where the thermal noise has been rescaled to $\eta_T=
\zeta_T/[3H(1+Q)]$ from Eq.~(\ref{eomWI}), after we take the slow-roll
approximation. 

In the case of the quantum noise, we can perform the two-point
correlation function for Eq.~(\ref{noise_qu}) in the slow-roll
approximation:

\begin{equation} \label{noise_qu_SRA}
\xi_q({\bf x},t) \approx - 3H\left(1 + Q
\right)\frac{\partial}{\partial t}\Phi_{<}({\bf x},t).  \;
\end{equation}
The correlation function for the quantum noise is given by Eq.~(2.12)
in Ref.~\cite{LAS2013} in the absence of a thermal bath. This
expression can be generalized for the case of warm inflation, which is
given by Eq.~(4.7) in Ref.~\cite{LAS2013}, although obtained  for a
different coarse graining  of the inflaton field.  {}From
Eq.~(\ref{noise_qu}), but in momenta space and expressing that
equation in the conformal time variable, $\tau=-[a(t)H]^{-1}$, we
obtain that

\begin{equation} \label{correl_J. Cosmol. Astropart. Phys.}
\langle \xi_q ({\bf k}, t) \xi_q ({\bf k}', t') \rangle = \delta({\bf
  k+k'})(\tau \tau')^2H^4(1+2n_k) \left[f_k(\tau)f_k^*(\tau')(1+n_k) +
  f_k^*(\tau)f_k(\tau') n_k\right]~.
\end{equation}
{}For the quantum noise in the slow-roll approximation,
Eq.~(\ref{noise_qu_SRA}), one obtains

\begin{equation} \label{fk_J. Cosmol. Astropart. Phys.}
f_k(\tau)= - \frac{3\left(1 + Q \right)}{\tau}\frac{\partial
  W(k,\tau)}{\partial \tau}\phi_k(\tau)  \; ,
\end{equation}
with

\begin{equation} \label{phik_J. Cosmol. Astropart. Phys.}
\phi_k(\tau)= \frac{H\sqrt{\pi}}{2}(|\tau|)^{3/2}
H_{\mu}^{(1)}(k|\tau|) \; ,
\end{equation}
where $H_{\mu}^{(1)}(k|\tau|)$ is the Hankel function of the first
kind and $\mu=\sqrt{9/4-3\eta}$, where $\eta$ is the slow-roll
coefficient given in Eq.~(\ref{sraparameters}).

Using the step filter function $W(k,\tau)=W(k+\epsilon\tau)$ and
performing the inverse space-Fourier transform of
Eq.~(\ref{correl_J. Cosmol. Astropart. Phys.}), one obtains that

\begin{equation} \label{correl_qu_NEW_FULL}
\langle \xi_q ({\bf x}, t) \xi_q ({\bf x}', t') \rangle =
\frac{H^3\epsilon^3}{16\pi}|H_{\mu}^{(1)}(\epsilon)|^2
\left(1+2n_{\tilde{k}}\right)
\frac{\sin[\epsilon a(t)H|{\bf x-x'}|]}{\epsilon a(t)H|{\bf x-x'}|}
\delta(t-t')~,
\end{equation}
where 

\begin{equation} \label{correl_nk}
n_{\tilde{k}}=\frac{1}{\exp{(\epsilon H/T)}-1}~.
\end{equation}
One particularly convenient choice for $\epsilon$ is
$\epsilon=1/(2\pi)$, which introduces the ratio $T_H/T$ in the
particle distribution, where $T_H=H/(2\pi)$ is the Gibbons-Hawking
temperature, and warm and cold inflation regimes can be naturally
defined in terms of $T_H$, $T> T_H$ and $T< T_H$, respectively.  {}For
this choice of $\epsilon$ and due to the fact that the slow-roll
coefficient $\eta$ is very small during inflation,  one can
approximate Eq.~(\ref{correl_qu_NEW_FULL}) to

\begin{equation} \label{correl_qu_NEW_FULL2}
\langle \xi_q ({\bf x}, t) \xi_q ({\bf x}', t') \rangle =
\frac{H^3}{4\pi^2}(1+2n_{\tilde{k}}) \frac{\sin[ a(t)T_H|{\bf
      x-x'}|]}{a(t)T_H|{\bf x-x'}|}   \delta(t-t')~.
\end{equation}

To obtain the Fokker-Planck diffusion coefficient, we need to rewrite
Eq.~(\ref{eom_appendix}) in the form of Eq.~(\ref{eom}). The
coefficients are obtained by multiplying the noises, whose correlation
function are given by $\delta(t-t')$ (with the proper normalizations
considered). In addition, we take  the limit of one worldline ${\bf
  x}={\bf x'}$. In this case, the quantum noise becomes simply

\begin{equation}\label{qn_appendix}
\eta_q = \frac{H^{3/2}}{2\pi}\sqrt{1+2n_{\tilde{k}}}\; \zeta_q~,
\end{equation}
and we obtain $\langle \zeta_q (t) \zeta_q (t') \rangle
=\delta(t-t')$.  On the other hand, the thermal correlation
Eq.~(\ref{both_correll_appendix}) involves a spatial Dirac-delta
function,  $\delta^3({\bf x} - {\bf x}')$, and a $a^{-3}$ factor.
Since we want the correlation function accumulated in one Hubble time,
$\Delta t\approx H^{-1}$, one obtains $a^{-3}=\exp{(-3H\Delta
  t)}\approx 20$. On the other hand,  one notice that $\delta^3({\bf
  x} - {\bf x}')$ corresponds to an inverse volume factor. The natural
volume to be taken is the de Sitter volume of the horizon, $V_H$,
which we obtain using the length scale $\approx H^{-1}$, and associate it with the
spatial Dirac delta, $ \delta({\bf x} - {\bf x}') \to 1/V_H =
1/(\frac{4 \pi}{3 H^3})$.  Therefore, one can approximate
the correlation function for the thermal noise $\eta_T$,
Eq.~(\ref{both_correll_appendix}), as

\begin{align}\label{thermal_correll_appendix}
\langle \eta_T (t) \eta_T (t') \rangle &= \frac{H^3}{4 \pi^2}\frac{
  \pi Q }{10(1+Q)^2}\left(\frac{T}{H}\right) \delta(t-t') \; .
\end{align}
{}From this result, we can rewrite

\begin{equation}\label{tn_appendix}
\eta_T = \frac{H^{3/2}}{2\pi}\sqrt{\frac{ \pi Q
  }{10(1+Q)^2}\left(\frac{T}{H}\right)}\; \zeta_T \;,
\end{equation}
and where $\langle \zeta_T (t) \zeta_T (t') \rangle =\delta(t-t')$.

{}From Eqs.~(\ref{eom_appendix}), (\ref{qn_appendix}) and
(\ref{tn_appendix}), one obtains

\begin{align}
\dot{\varphi}&=-\frac{V_{,\varphi}}{3H(1+Q)} +
\frac{H^{3/2}}{2\pi}\sqrt{1+2n_k} \zeta_q +
\frac{H^{3/2}}{2\pi}\sqrt{\left(\frac{T}{H}\right)\frac{\pi
    Q}{10(1+Q)^2}} \zeta_T \; ,
\end{align}
which, compared to Eq.~(\ref{eom}), finally gives

\begin{align}\label{coeffs_WI_appendix}
D^{(2)}   = D^{(2)}_{\rm{(vac)}} + D^{(2)}_{\rm{(diss)}}  =
\frac{H^3}{8\pi^2}  \left[1 + 2 n_k + \frac{\pi Q
  }{10(1+Q)^2}\left(\frac{T}{H}\right)\right]~.
\end{align}

%%%%%%%%%%%%%%%%%%%%%%%%%%%%%%%%%%%%%%%%%%%%%%%%%%%%%%%%%%%%%%%%%%%%%%%%%%%%%

\section{Numerical Analysis} \label{numerical_analysis}

In order to perform the numerical analysis, we have integrated the
background equations along with the {}Fokker-Planck coefficients. The
background equations of warm inflation in the slow-roll approximation
(SRA), Eqs.~(\ref{eoms_sra}) and (\ref{eoms_sra2}), can be suitably
rewritten in terms of the number of {\it e}-folds $N_e$ as

\begin{align}
\frac{d\phi/M_P}{dN_e} &= -
\left(\frac{\phi}{M_P}\right)\frac{\kappa}{(1+Q)} \;
, \label{dphidN}\\ \frac{d\ln Q}{dN_e} &=  \frac{1}{(1+7Q)}(10\epsilon
- 6\eta + 8\kappa) \; , \label{dQdN}\\ \frac{d\ln (T/H)}{dN_e} &=
\frac{2}{(1+7Q)}\left(\frac{2+4Q}{1+Q}\epsilon - \eta +
\frac{1-Q}{1+Q}\kappa\right)\; \label{dnudN},
\end{align}
where
$\kappa =M_P^2 \left(\frac{V_{,\phi}/\phi}{V}\right)$.
{}From these equations, the second and third ones are given
specifically for the dissipation term $\Upsilon$ considered in this
work, Eq.~({\ref{upsilon}}). 

These SRA equations for the chaotic potential, Eq.~(\ref{potential}),
are given by 

\begin{align}
\frac{d \phi/M_P}{dN_e} &=
-\frac{2n}{1+Q}\left(\frac{\phi}{M_P}\right)^{-1} \;
, \label{dphidN_chaotic}\\  \frac{d\ln Q}{dN_e}     &=
\frac{4n(7-n)}{1+7Q} \left(\frac{\phi}{M_P}\right)^{-2} \;
, \label{dQdN_chaotic}\\ \frac{d\ln (T/H)}{dN_e}   &=
\frac{8n(1+nQ)}{(1+Q)(1+7Q)}\left(\frac{\phi}{M_P}\right)^{-2} \;
, \label{dvudN_chaotic}
\end{align}
while for the hilltop potential, Eq.~(\ref{potential_hybrid_nneq0}),
these are given by

\begin{align}
\frac{d\ln \phi/M_P}{dN_e} &=
\frac{\left|\gamma\right|}{1+Q}
\frac{\left(\frac{\phi}{M_P}\right)^{2n-2}}
{1-\frac{\left|\gamma\right|}{2n}\left(\frac{\phi}{M_P}\right)^{2n}}
\; ,\label{dphidN_hilltop}\\ \frac{d\ln Q}{dN_e} &=-
\frac{\gamma}{1+7Q}
\frac{\left(\frac{\phi}{M_P}\right)^{2n-2}}
{1-\frac{\left|\gamma\right|}{2n}\left(\frac{\phi}{M_P}\right)^{2n}}
\left[14 -12n -\frac{5\left|\gamma\right|
    \left(\frac{\phi}{M_P}\right)^{2n}}
{1-\frac{\left|\gamma\right|}{2n}\left(\frac{\phi}{M_P}\right)^{2n}}
  \right] \; ,\label{dQdN_hilltop} \\ \frac{d\ln (T/H)}{dN_e} &=-
\frac{2\left|\gamma\right|}{1+7Q}
\frac{\left(\frac{\phi}{M_P}\right)^{2n-2}}
{1-\frac{\left|\gamma\right|}{2n}\left(\frac{\phi}{M_P}\right)^{2n}}
\left[\frac{2}{1+Q} - 2n - \frac{1+2Q}{1+Q}
  \frac{\left|\gamma\right|\left(\frac{\phi}{M_P}\right)^{2n}}
{1-\frac{\left|\gamma\right|}{2n}\left(\frac{\phi}{M_P}\right)^{2n}}
  \right] \; .\label{dvudN_hilltop}
\end{align}

In terms of the dimensionless variables,

\begin{eqnarray}
L&=&v,\\ Q&=&\Upsilon'/3L,\\ \epsilon&=&\frac{1}{2}\left(V_{,x}/V\right)^2,
\\ \eta&=&V_{,xx}/V,\\ \kappa&=&(V_{,x}/x)/V~,
\end{eqnarray}
the dimensionless versions of the SRA equations presented above keep
their forms, except for the identification of the dimensionless inflaton
field $x=\phi/M_P$. In terms of these variables, the dimensionless
{}Fokker-Planck coefficients $D^{(1)}$ and $D^{(2)}$,
Eqs.~(\ref{arraste}) and (\ref{diffusion}), are given, respectively,
by

\begin{eqnarray} 
\label{dimd1}
d^{(1)} &=& -\frac{v_{,x}}{2nL(1+Q)},  \\
\label{dimd2}
d^{(2)} &=& \frac{\lambda}{12n^2}  \frac{L^{3}}{8\pi^2 } \left[   1 +
  \frac{2}{e^{L/T'} - 1} + \left(\frac{T'}{L}\right) \frac{\pi
    Q}{10(1+Q)^2} \right],
\end{eqnarray}
where

\begin{eqnarray}
\label{relationd1}
D^{(1)} &=& \sqrt{\frac{\lambda}{6n}} M_p^2 d^{(1)},\\
\label{relationd2}
D^{(2)} &=& \sqrt{\frac{2n\lambda}{3}} M_p^3 d^{(2)}. 
\end{eqnarray}

The FDR condition given in the main text, Eq.~(\ref{condition}), in
terms of the dimensionless variables becomes

\begin{eqnarray}\label{condition_dimless}
\frac{v'(x)}{L^2(1+Q)} \ll \sqrt{\frac{\lambda}{6n}}  \frac{L}{2\pi }
\left[   1 + \frac{2}{e^{L/T'} - 1} + \left(\frac{T'}{L}\right)
  \frac{\pi Q}{10(1+Q)^2} \right]^{1/2}.
\end{eqnarray}
The FDR ends when this equation becomes an equality, which provides us
with the value $x=x_{\rm{th}}$ for which this regime ends. This is the
dimensionless version of $\varphi=\varphi_{\rm{th}}$, $x=x_{\rm{th}}$
that we have used in our results.

We have analyzed eternal inflation with the concomitant study of
$\varphi_{\rm{th}}$ and $N_{\rm{th}}$ when varying the dissipation
ratio $Q_*$.  This has been done by integrating the background SRA
equations backwards from  the end of inflation (given by the slow-roll
parameters) for our chosen  $Q_*$ interval (by choosing a suitable
$Q_f$ value).  Since eternal inflation occurs from the beginning of
inflation until  some $x=x_{\rm{th}}$, we perform a backwards loop on
the number of  {\it e}-folds for Eq.~(\ref{condition_dimless}) from $x_f$
(where the FRD condition  is not satisfied) until the FDR condition
becomes an equality, obtaining $x=x_{\rm{th}}$. {}Finally, the value
of the number of {\it e}-folds at $x=x_{\rm{th}}$ gives us $N_{\rm{th}}$.
This procedure is repeated  for each value of $Q_*$  and we obtain the
functional relations  $\varphi=\varphi_{\rm{th}}(Q_*)$, and
$N=N_{\rm{th}}(Q_*)$.

%%%%%%%%%%%%%%%%%%%%%%%%%%%%%%%%%%%%%%%%%%%%%%%%%%%%

\acknowledgments G.S.V was supported by Coordena\c{c}\~ao de Aperfei\c{c}oamento de
Pessoal de N\'{\i}vel Superior (CAPES), L.A.S. was supported by Funda\c{c}\~ao de Amparo \`a
pesquisa do Estado de S\~ao Paulo (FAPESP) and R.O.R is partially
supported by research grants from Conselho Nacional de Desenvolvimento
Cient\'{\i}fico e Tecnol\'ogico (CNPq)  and Funda\c{c}\~ao Carlos
Chagas Filho de Amparo \`a Pesquisa do Estado do Rio de Janeiro
(FAPERJ).

%%%%%%%%%%%%%%%%%%%%%%%%%%%%%%%%%%%%%%%%%%%%%%%%%%%%%%%%%%%%%%%%%%%%%%%%%


\begin{thebibliography}{99}

\bibitem{Guth2007}
 A.~H.~Guth,
  %``Eternal inflation and its implications,''
  J.\ Phys.\ A {\bf 40}, 6811 (2007).
%  [hep-th/0702178 [HEP-TH]].
  %%CITATION = HEP-TH/0702178;%%
  %149 citations counted in INSPIRE as of 29 Jul 2015

\bibitem{winitzki2009eternal}
S.~Winitzki, 
{\it Eternal Inflation}, 
(World Scientific Publishing,
 Singapore, 2009).

\bibitem{guth2000inflation}
 A.~H.~Guth,
  %``Inflation and eternal inflation,''
  Phys.\ Rep.\  {\bf 333}, 555 (2000).
%  [astro-ph/0002156].
  %%CITATION = ASTRO-PH/0002156;%%
  %190 citations counted in INSPIRE as of 29 juil. 2015

\bibitem{Vilenkin:1983xq} 
  A.~Vilenkin,
  %``The Birth of Inflationary Universes,''
  Phys.\ Rev.\ D {\bf 27}, 2848 (1983).
  %%CITATION = PHRVA,D27,2848;%%

\bibitem{Guth:1985ya} 
  A.~H.~Guth and S.~Y.~Pi,
  %``The Quantum Mechanics of the Scalar Field in the New Inflationary Universe,''
  Phys.\ Rev.\ D {\bf 32}, 1899 (1985).
  %%CITATION = PHRVA,D32,1899;%%

\bibitem{Linde:1986fc} 
  A.~D.~Linde,
  %``Eternal Chaotic Inflation,''
  Mod.\ Phys.\ Lett.\  A {\bf 1}, 81 (1986).
  %%CITATION = MPLAE,A1,81;%%

\bibitem{Linde:1986fd} 
  A.~D.~Linde,
  %``Eternally Existing Selfreproducing Chaotic Inflationary Universe,''
  Phys.\ Lett.\ B {\bf 175}, 395 (1986).
  %%CITATION = PHLTA,B175,395;%%

\bibitem{vilenkin2007measure}
A.~Vilenkin,
  %``A Measure of the multiverse,''
  J.\ Phys.\ A {\bf 40}, 6777 (2007).
%  [hep-th/0609193].
  %%CITATION = HEP-TH/0609193;%%
  %68 citations counted in INSPIRE as of 29 Jul 2015

\bibitem{garriga2006probabilities}
 J.~Garriga, D.~Schwartz-Perlov, A.~Vilenkin and S.~Winitzki,
  %``Probabilities in the inflationary multiverse,''
  J. Cosmol. Astropart. Phys. {\bf 01} (2006) 017.
%  [hep-th/0509184].
  %%CITATION = HEP-TH/0509184;%%
  %142 citations counted in INSPIRE as of 29 juil. 2015


\bibitem{zhang2015testing}
P.~Zhang and M.~C.~Johnson,
  %``Testing eternal inflation with the kinetic Sunyaev Zel'dovich effect,''
  J. Cosmol. Astropart. Phys. {\bf 06}, (2015) 046.
%  [arXiv:1501.00511 [astro-ph.CO]].
  %%CITATION = ARXIV:1501.00511;%%
  %1 citations counted in INSPIRE as of 29 juil. 2015

\bibitem{wainwright2014simulating}
C.~L.~Wainwright, M.~C.~Johnson, H.~V.~Peiris, A.~Aguirre, L.~Lehner and S.~L.~Liebling,
  %``Simulating the universe(s): from cosmic bubble collisions to cosmological observables with numerical relativity,''
  J. Cosmol. Astropart. Phys. {\bf 1403}, 030 (2014).
%  [arXiv:1312.1357 [hep-th]].
  %%CITATION = ARXIV:1312.1357;%%
  %12 citations counted in INSPIRE as of 29 juil. 2015

\bibitem{wainwright2014simulating2}
 C.~L.~Wainwright, M.~C.~Johnson, A.~Aguirre and H.~V.~Peiris,
  %``Simulating the universe(s) II: phenomenology of cosmic bubble collisions in full General Relativity,''
  J. Cosmol. Astropart. Phys. {\bf 1410}, 024 (2014).
%  [arXiv:1407.2950 [hep-th]].
  %%CITATION = ARXIV:1407.2950;%%
  %8 citations counted in INSPIRE as of 29 juil. 2015

\bibitem{feeney2011first}
S.~M.~Feeney, M.~C.~Johnson, D.~J.~Mortlock and H.~V.~Peiris,
  %``First Observational Tests of Eternal Inflation,''
  Phys.\ Rev.\ Lett.\  {\bf 107}, 071301 (2011).
%  [arXiv:1012.1995 [astro-ph.CO]].
  %%CITATION = ARXIV:1012.1995;%%
  %33 citations counted in INSPIRE as of 29 Jul 2015

\bibitem{feeney2011first2}
S.~M.~Feeney, M.~C.~Johnson, D.~J.~Mortlock and H.~V.~Peiris,
  %``First Observational Tests of Eternal Inflation: Analysis Methods and WMAP 7-Year Results,''
  Phys.\ Rev.\ D {\bf 84}, 043507 (2011).
%  [arXiv:1012.3667 [astro-ph.CO]].
  %%CITATION = ARXIV:1012.3667;%%
  %37 citations counted in INSPIRE as of 29 Jul 2015

\bibitem{aguirre2011status}
 A.~Aguirre and M.~C.~Johnson,
  %``A Status report on the observability of cosmic bubble collisions,''
  Rept.\ Prog.\ Phys.\  {\bf 74}, 074901 (2011).
%  [arXiv:0908.4105 [hep-th]].
  %%CITATION = ARXIV:0908.4105;%%
  %28 citations counted in INSPIRE as of 29 Jul 2015

\bibitem{smolyaninov2013experimental}
 I.~I.~Smolyaninov, B.~Yost, E.~Bates and V.~N.~Smolyaninova,
  %``Experimental demonstration of metamaterial multiverse in a ferrofluid,''
  Opt.\ Express {\bf 21}, 14918 (2013).
%  [arXiv:1301.6055 [physics.optics]].
  %%CITATION = ARXIV:1301.6055;%%
  %4 citations counted in INSPIRE as of 29 Jul 2015
  
\bibitem{mukhanov2014inflation}
  V.~Mukhanov,
  %``Inflation without Selfreproduction,''
  Fortsch.\ Phys.\  {\bf 63}, 36 (2015).
%  [arXiv:1409.2335 [astro-ph.CO]].
  %%CITATION = ARXIV:1409.2335;%%
  %11 citations counted in INSPIRE as of 29 Jul 2015

\bibitem{kinney2014negative}
 W.~H.~Kinney and K.~Freese,
  %``Negative running can prevent eternal inflation,''
  J. Cosmol. Astropart. Phys. {\bf 1501}, 040 (2015).
%  [arXiv:1404.4614 [astro-ph.CO]].
  %%CITATION = ARXIV:1404.4614;%%
  %7 citations counted in INSPIRE as of 29 Jul 2015


\bibitem{Planck2013}  P.~A.~R.~Ade {\it et al.} [Planck Collaboration],
  %``Planck 2013 results. XXII. Constraints on inflation,''
  Astron.\ Astrophys.\  {\bf 571}, A22 (2014).
%  [arXiv:1303.5082 [astro-ph.CO]].
  %%CITATION = ARXIV:1303.5082;%%

\bibitem{BICEP2}
  P.~A.~R.~Ade {\it et al.} [BICEP2 Collaboration],
  %``Detection of $B$-Mode Polarization at Degree Angular Scales by BICEP2,''
  Phys.\ Rev.\ Lett.\  {\bf 112}, 241101 (2014).
%  [arXiv:1403.3985 [astro-ph.CO]].
  %%CITATION = ARXIV:1403.3985;%%
  %1092 citations counted in INSPIRE as of 29 Jul 2015

\bibitem{brandenberger2015can}
R.~Brandenberger, R.~Costa and G.~Franzmann,
  %``Can Back-Reaction Prevent Eternal Inflation?,''
  arXiv:1504.00867 [hep-th].
  %%CITATION = ARXIV:1504.00867;%%
  %1 citations counted in INSPIRE as of 29 Jul 2015

\bibitem{Berera:1995ie} 
  A.~Berera,
  %``Warm inflation,''
  Phys.\ Rev.\ Lett.\  {\bf 75}, 3218 (1995).
%  [astro-ph/9509049].
  %%CITATION = ASTRO-PH/9509049;%%

\bibitem{lyth2009primordial}
D.~Lyth and A.~Liddle, 
{\it The Primordial Density Perturbation: Cosmology, Inflation and the Origin of Structure}. 
Cambridge University Press, (2009).


\bibitem{WIreviews1}
  A.~Berera, I.~G.~Moss and R.~O.~Ramos,
  %``Warm Inflation and its Microphysical Basis,''
  Rept.\ Prog.\ Phys.\  {\bf 72}, 026901 (2009).
%  [arXiv:0808.1855 [hep-ph]].
  %%CITATION = ARXIV:0808.1855;%%
  %90 citations counted in INSPIRE as of 29 Jul 2015

\bibitem{WIreviews2}
M.~Bastero-Gil and A.~Berera,
  %``Warm inflation model building,''
  Int.\ J.\ Mod.\ Phys.\ A {\bf 24}, 2207 (2009).
%  [arXiv:0902.0521 [hep-ph]].
  %%CITATION = ARXIV:0902.0521;%%
  %56 citations counted in INSPIRE as of 29 Jul 2015

\bibitem{warmpert1}
A.~N.~Taylor and A.~Berera,
  %``Perturbation spectra in the warm inflationary scenario,''
  Phys.\ Rev.\ D {\bf 62}, 083517 (2000).
%  [astro-ph/0006077].
  %%CITATION = ASTRO-PH/0006077;%%
  %85 citations counted in INSPIRE as of 29 Jul 2015

\bibitem{warmpert2}
 L.~M.~H.~Hall, I.~G.~Moss and A.~Berera,
  %``Scalar perturbation spectra from warm inflation,''
  Phys.\ Rev.\ D {\bf 69}, 083525 (2004).
%  [astro-ph/0305015].
  %%CITATION = ASTRO-PH/0305015;%%
  %103 citations counted in INSPIRE as of 29 Jul 2015

\bibitem{ng1}
  C.~-H.~Wu, K.~-W.~Ng, W.~Lee, D.~-S.~Lee and Y.~-Y.~Charng,
%{\it Quantum noise and a low cosmic microwave background quadrupole},
  J. Cosmol. Astropart. Phys. {\bf 0702}, 006  (2007).
%  [astro-ph/0604292].
  %%CITATION = ASTRO-PH/0604292;%%

\bibitem{ng2}   W.~Lee, K.~-W.~Ng, I-C.~Wang and C.~-H.~Wu,
{\it Trapping effects on inflation},
  Phys.\ Rev.\ D {\bf 84}, 063527  (2011).
%  [arXiv:1101.4493 [hep-th]].
  %%CITATION = ARXIV:1101.4493;%%

\bibitem{Liu:2014ifa}   G.~C.~Liu, K.~W.~Ng and I.~C.~Wang,
  %``Naturally large tensor-to-scalar ratio in inflation,''
  Phys.\ Rev.\ D {\bf 90}, 103531 (2014).
%  [arXiv:1409.3661 [hep-ph]].
  %%CITATION = ARXIV:1409.3661;%%

\bibitem{LAS2013}
R.~O.~Ramos and L.~A.~da Silva,
  %``Power spectrum for inflation models with quantum and thermal noises,''
  J. Cosmol. Astropart. Phys. {\bf 1303}, 032 (2013).
%  [arXiv:1302.3544 [astro-ph.CO]].
  %%CITATION = ARXIV:1302.3544;%%
  %13 citations counted in INSPIRE as of 29 Jul 2015

\bibitem{bartrum2014importance}
S.~Bartrum, M.~Bastero-Gil, A.~Berera, R.~Cerezo, R.~O.~Ramos and J.~G.~Rosa,
  %``The importance of being warm (during inflation),''
  Phys.\ Lett.\ B {\bf 732}, 116 (2014).
%  [arXiv:1307.5868 [hep-ph]].
  %%CITATION = ARXIV:1307.5868;%%
  %29 citations counted in INSPIRE as of 29 juil. 2015

\bibitem{bastero2014cosmological}
  M.~Bastero-Gil, A.~Berera, I.~G.~Moss and R.~O.~Ramos,
  %``Cosmological fluctuations of a random field and radiation fluid,''
  J. Cosmol. Astropart. Phys. {\bf 1405}, 004 (2014).
%  [arXiv:1401.1149 [astro-ph.CO]].
  %%CITATION = ARXIV:1401.1149;%%
  %13 citations counted in INSPIRE as of 29 juil. 2015

\bibitem{Bastero-Gil:2014raa} 
  M.~Bastero-Gil, A.~Berera, I.~G.~Moss and R.~O.~Ramos,
  %``Theory of non-Gaussianity in warm inflation,''
  J. Cosmol. Astropart. Phys. {\bf 1412}, 008 (2014)
%  [arXiv:1408.4391 [astro-ph.CO]].
  %%CITATION = ARXIV:1408.4391;%%

\bibitem{Ijjas:2013vea} 
  A.~Ijjas, P.~J.~Steinhardt and A.~Loeb,
  %``Inflationary paradigm in trouble after Planck2013,''
  Phys.\ Lett.\ B {\bf 723}, 261 (2013).
%  [arXiv:1304.2785 [astro-ph.CO]].
  %%CITATION = ARXIV:1304.2785;%%

\bibitem{Guth:2013sya} 
  A.~H.~Guth, D.~I.~Kaiser and Y.~Nomura,
  %``Inflationary paradigm after Planck 2013,''
  Phys.\ Lett.\ B {\bf 733}, 112 (2014).
%  [arXiv:1312.7619 [astro-ph.CO]].
  %%CITATION = ARXIV:1312.7619;%%

\bibitem{Linde:2014nna} 
  A.~Linde,
  %``Inflationary Cosmology after Planck 2013,''
  arXiv:1402.0526 [hep-th].
  %%CITATION = ARXIV:1402.0526;%%

\bibitem{original}
A.~A.~Starobinsky,
  %``Stochastic De Sitter (inflationary) Stage In The Early Universe,''
  Lect.\ Notes Phys.\  {\bf 246}, 107 (1986).
  %%CITATION = LNPHA,246,107;%%
  %64 citations counted in INSPIRE as of 29 Jul 2015

\bibitem{Starobinsky1994}
A.~A.~Starobinsky and J.~Yokoyama,
  %``Equilibrium state of a selfinteracting scalar field in the De Sitter background,''
  Phys.\ Rev.\ D {\bf 50}, 6357 (1994).
%  [astro-ph/9407016].
  %%CITATION = ASTRO-PH/9407016;%%
  %245 citations counted in INSPIRE as of 29 Jul 2015


\bibitem{PhysRevD.26.1231}
A.~Vilenkin and L.~H.~Ford,
  %``Gravitational Effects upon Cosmological Phase Transitions,''
  Phys.\ Rev.\ D {\bf 26}, 1231 (1982).
  %%CITATION = PHRVA,D26,1231;%%
  %515 citations counted in INSPIRE as of 29 Jul 2015

\bibitem{Linde1982335}
A.~D.~Linde,
  %``Scalar Field Fluctuations in Expanding Universe and the New Inflationary Universe Scenario,''
  Phys.\ Lett.\ B {\bf 116}, 335 (1982).
  %%CITATION = PHLTA,B116,335;%%
  %598 citations counted in INSPIRE as of 29 Jul 2015

\bibitem{Starobinsky1982175}
A.~A.~Starobinsky,
  %``Dynamics of Phase Transition in the New Inflationary Universe Scenario and Generation of Perturbations,''
  Phys.\ Lett.\ B {\bf 117}, 175 (1982).
  %%CITATION = PHLTA,B117,175;%%
  %1703 citations counted in INSPIRE as of 29 Jul 2015

\bibitem{Winitzki1996}
S.~Winitzki and A.~Vilenkin,
  %``Uncertainties of predictions in models of eternal inflation,''
  Phys.\ Rev.\ D {\bf 53}, 4298 (1996).
%  [gr-qc/9510054].
  %%CITATION = GR-QC/9510054;%%
  %46 citations counted in INSPIRE as of 29 Jul 2015

\bibitem{madureira1996giant}
A.~J.~R.~Madureira, P.~H\"anggi, H.~S.~Wio, 
  %``Giant suppression of the activation rate in the presence of correlated white noise sources,''
  Phys.\ Lett.\ A {\bf 217}, 248 (1996).

\bibitem{Berera:2007qm} 
  A.~Berera, I.~G.~Moss and R.~O.~Ramos,
  %``Local Approximations for Effective Scalar Field Equations of Motion,''
  Phys.\ Rev.\ D {\bf 76}, 083520 (2007).
%  [arXiv:0706.2793 [hep-ph]].
  %%CITATION = ARXIV:0706.2793;%%

\bibitem{Bastero-Gil2011}
M.~Bastero-Gil, A.~Berera and R.~O.~Ramos,
  %``Dissipation coefficients from scalar and fermion quantum field interactions,''
  J. Cosmol. Astropart. Phys. {\bf 1109}, 033 (2011).
%  [arXiv:1008.1929 [hep-ph]].
  %%CITATION = ARXIV:1008.1929;%%
  %53 citations counted in INSPIRE as of 29 Jul 2015

\bibitem{BasteroGil:2012cm} 
  M.~Bastero-Gil, A.~Berera, R.~O.~Ramos and J.~G.~Rosa,
  %``General dissipation coefficient in low-temperature warm inflation,''
  J. Cosmol. Astropart. Phys. {\bf 1301}, 016 (2013).
%  [arXiv:1207.0445 [hep-ph]].
  %%CITATION = ARXIV:1207.0445;%%

\bibitem{boubekeur2005hilltop}
L.~Boubekeur and D.~H.~Lyth,
  %``Hilltop inflation,''
  J. Cosmol. Astropart. Phys. {\bf 0507}, 010 (2005).
%  [hep-ph/0502047].
  %%CITATION = HEP-PH/0502047;%%
  %176 citations counted in INSPIRE as of 29 juil. 2015

\bibitem{Planck2015}   P.~A.~R.~Ade {\it et al.} [Planck Collaboration],
  %``Planck 2015 results. XX. Constraints on inflation,''
  arXiv:1502.02114 [astro-ph.CO].
  %%CITATION = ARXIV:1502.02114;%%

\end{thebibliography}
\end{document}